\documentclass[authoryear]{elsarticle} 

\usepackage{fullpage}
\journal{arXiv}

\usepackage{enumitem}

\usepackage{upgreek}
\usepackage{mathtools}
\usepackage{rotating}
\usepackage{amsmath}
\usepackage{amssymb}
\usepackage{bbm}
\usepackage{subcaption}
\usepackage{graphicx}
\usepackage{url}
\usepackage[linesnumbered]{algorithm2e}
\usepackage{placeins}

\def\bSig\mathbf{\Sigma}

\newcommand {\pbi}{\begin{itemize}}
\newcommand {\pei}{\end{itemize}}

\newcommand {\pbc}{\begin{center}}
\newcommand {\pec}{\end{center}}
\newcommand {\pbe}{\begin{eqnarray*}}
\newcommand {\pee}{\end{eqnarray*}}
\newcommand {\pben}{\begin{eqnarray}}
\newcommand {\peen}{\end{eqnarray}}
\let\hat=\widehat
\let\geq=\geqslant

\usepackage {tikz}
\usetikzlibrary{arrows}
\begin{document}
	
\begin{frontmatter}
		
\title{Uncertainties in estimating the effect of climate change on 100-year return value for significant wave height}
\author[mos,melbourne]{Kevin Ewans}
\author[lancs,shelluk]{Philip Jonathan\corref{cor1}}
\address[mos]{MetOcean Research Ltd, New Plymouth 4310, New Zealand.}	
\address[melbourne]{Department of Infrastructure Engineering, University of Melbourne, Melbourne, VIC 3010, Australia.}
\address[lancs]{Department of Mathematics and Statistics, Lancaster University LA1 4YF, United Kingdom.}		
\address[shelluk]{Shell Research Limited, London SE1 7NA, United Kingdom.}
\cortext[cor1]{Corresponding author {\tt philip.jonathan@shell.com}}

\begin{abstract}
The process of estimating the effect of a changing climate on the severity of future ocean storms is plagued by large uncertainties; for safe design and operation of offshore structures, it is nevertheless important that best possible estimates of climate effects is made given the available data. We explore the variability in estimates of 100-year return value of significant wave height ($H_S$), and changes in estimates over a period of time, for output of WAVEWATCH-III models from 7 representative CMIP5 GCMs, and the FIO-ESM v2.0 CMIP6 GCM. Non-stationary extreme value analysis of peaks-over-threshold (POT) and block maxima (BM) using Bayesian inference provide posterior estimates of return values as a function of time; MATLAB software for the extreme value analysis is provided. Best overall estimates for return values, and changes in return value over the period 1979-2100, are calculated by averaging estimates for individual GCMs. We focus attention on neighbourhoods of locations east of Madagascar (EoM) and south of Australia (SoA) where a previous study of CMIP5 output by \cite{MccEA20} reported significant decrease and increase in $H_S$ respectively, under RCP4.5 and RCP8.5 climate scenarios. There is large variation between return value estimates from different GCMs, and with longitude and latitude within each neighbourhood for estimates based on samples corresponding to $\le$ 165 years of model output; these sources of uncertainty tend to be larger than that due to typical modelling choices (such as choice of threshold for POT, or block length for BM). However, we also find that careful threshold choice and block length are critical EoM, because of the presence of a mixed population of storms there. Nevertheless, there is general evidence supporting the trends reported by \cite{MccEA20}, but these findings are conditional on the choice of 8 GCMs being representative of climate evolution. We use simple randomisation testing to identify ``significant'' departures from steady climate. The long 700-year pre-industrial control (piControl) output of the CMIP6 GCM offers an excellent opportunity to quantify the apparent inherent variability in return value as a function of time, estimated using a subsample of output corresponding to a continuous time interval of between 20 and 160 years in length, where no climate forcing is present. We find large variation in return value estimates of approximately $\pm 15\%$ made from samples corresponding to periods of time of around 50 years drawn from piControl data. 
\end{abstract}


\end{frontmatter}

\section{Introduction} \label{Sct:Int}
%
The effect of a changing climate on the occurrence and intensity of ocean storms is receiving considerable attention, but efforts to identify and quantify effects and to attribute these to climate change remain hampered by large uncertainties. There is agreement on some climate influence, such as the increasing proportion of intense tropical cyclones (Category 4–5) and peak wind speeds of the most intense tropical cyclones globally with increasing global warming (\citealt{IPCC-SPM}). Nevertheless, predictions of possible changes in the future ocean wave climate and particularly sea state extremes are invariably very uncertain. Regardless, it is still important for safe design and execution of offshore and coastal activities that predictions of possible climate-related changes and their significance are made. In particular, climate change introduces additional uncertainty on the robustness of future design criteria.

It is now generally accepted that the historical data sets upon which design criteria are based are not stationary, as implicitly assumed. Inter-annual variability and also longer-term, aperiodic, multi-decadal atmospheric oscillations, result in a time-varying climate. Accordingly, even met-ocean databases spanning several decades are not stationary, due to these aperiodicities. There is also a mounting body of evidence that global warming since the onset of the industrial revolution is resulting in changes in the intensity of historical storms and wave climate (e.g. \citealt{YngRbl19}), and climate change projections out to the end of the 21st century indicate likely changes in the wind and wave climate, including their extremes (e.g., \citealt{MccEA20}, \citealt{MccEA22}). The consequence of a non-stationary climate is that met-ocean statistics change with time, albeit slowly if based on a long data set, and in particular, the 100-year return value for significant wave height ($H_S$) values estimated from data covering different decades will be different. In addition, future climate change trends can be expected to accentuate the temporal change in these statistics. Accordingly, better understanding the temporal effects and their uncertainties is needed to evaluate the robustness of estimates against future climate variability.

Estimates of extreme sea states for a future climate can be made from the output of General Circulation Models (GCMs), which have typically been run at relatively coarse spatial grid sizes and temporal resolutions and have not included, until recently, estimates of sea state parameters. Thus, predictions of wave fields from GCMs involve downscaling, either statistically or dynamically, from the coarse GCM grid parameter fields, usually sea level pressure or wind fields, associated with various climate change scenarios. A discussion of recent studies on predictions of likely changes in the wave climate based on these types of analyses of these types and the uncertainties involved can be found in \cite{EwnJnt20}. Most studies have been based on a global or regional analysis and have provided interesting perspectives of possible changes in the wave climate on those scales. \cite{MccEA20} performs a global analysis of the effect of climate change on $H_S$, combining standardised data from an ensemble of wave models forced by independent CMIP5 climate simulations under Representative Concentration Pathway (RCP) scenarios RCP4.5 and RCP8.5. They find for example that the magnitude of the 100-year $H_S$ event increases by 5 to 15\% over the Southern Ocean by the end of the 21st century, compared to the 1979–2005 period. \cite{MccEA22} extends the basis of the analysis of \cite{MccEA20} to include four 140-year wind-wave climate simulations forced with surface wind speed and sea ice concentration from two CMIP6 GCMs under Shared Socioeconomic Pathway (SSP) scenarios SSP126 and SSP585. They emphasise the advantages of a long 140-year interval of data to establish the presence and size of climate trends. The global COWCLIP2.0 ensemble of CMIP5-driven projections for ocean wave climate reported by \cite{MrmEA20} a valuable resource for broad scale coastal and offshore design. \cite{LbtEA21} reports predicted increases in $H_S$ of around 2m in the Southern Ocean for CMIP5 RCP8.5, whereas the north Pacific shows largest reductions of around 1.5m. \cite{SrdEA22} use a multi-model ensemble to study projected changes in extreme wave height indices (e.g. rough wave days) from COWCLIP2.0 under RCP4.5 and RCP8.5 emission scenarios for the Indian Ocean, and find some substantial spatial changes in time, including evidence for teleconnections. \cite{AlbEA22} uses COWCLIP ensemble members downscaled using the SWAN wave model to assess changes in New Zealand's wave climate, finding a general increase in wave height along the southern and western coasts. \cite{CssPrtEA2022} assesses the effects of inherent climate variability on trends in $H_S$ using a single model initial-condition large 100-member ensemble simulation for the period 1951–2010, and compares findings with those from other sources including wave reanalyses. Trends arising from internal climate variability are comparable in size to those caused by other factors, such as climate model uncertainty. They emphasise that at least 10 ensemble members are generally needed to detect trends found in the 100-member ensemble with confidence.

Detailed engineering design criteria are required for developments at a specific site, which involves a more focused analysis. Our objective in this paper is to undertake analyses to gain an appreciation of the uncertainties involved in estimating site-specific sea state extremes in a non-stationary climate, including those associated with estimating changes in the extremes in a possible future wave climate, and so also provide a perspective on the limitations in assessing how robust site-specific criteria derived from historical data might be. Specifically, how much confidence can we place on predictions of 100-year $H_S$, made with GCM-derived output data for projected future climates?

Dynamical downscaling involves running numerical wave prediction models using the GCM data as boundary conditions. This approach is computationally expensive, but provides a more direct estimate of the wave field and sea-state parameters such as $H_S$ and associated quantities. $H_S$ is the most common parameter used to describe the severity of a sea state, and the 100-year $H_S$ is the most common extremal parameter used to describe wave climate extremes; the 100-year $H_S$ is also the most common parameter found in offshore and coastal engineering design criteria. Accordingly, we focus our attention on estimating 100-year $H_S$ from data sets produced from running numerical wave models forced by GCMs. As our extreme value (EV) analyses include modelling of peaks over thresholds, a necessary requirement of the data sets is that they have sufficient temporal resolution for individual storms, and hence storm peak $H_S$ values to be determined. This is a significant constraint, limiting our study to a relatively small number of suitable data sets. Our study is based on eight wave data sets, seven of which are forced by independent CMIP5 (Coupled Model Intercomparison Project phase 5) GCMs, with a 6-hourly time step, and one forced by a CMIP6 GCM, with a 3-hourly time step. The CMIP5 data used here for each model is a subset of the data used by \cite{MccEA20} in their global study, and the CMIP6 data used here is a subset of that described by \cite{BaoEA20}. The data are introduced in Section~\ref{Sct:Dat} below. The CMIP5 output includes an historical period and future projections for two climate forcing scenarios, namely RCP4.5 and RCP8.5. The CMIP6 data includes a quasi-equilibrium 700-year pre-industrial (piControl) data set, a historical data set, three future scenario data sets - SSP126, SSP245, and SSP585, and two climate sensitive data sets. 

We focus on sets of locations in two regions, one off the east coast of Madagascar (henceforth ``EoM'') in the Indian Ocean (centred around 15$^\circ$S, 56$^\circ$E), and one south of Australia (``SoA'') in the Southern Ocean (centred around 48$^\circ$S, 142$^\circ$E). The two regions were selected as they correspond to significant decrease and increase in 100-year $H_S$ in the analysis of \cite{MccEA20}. Our interest is to quantify the statistical characteristics of 100-year $H_S$ at specific sites within the two regions. We expect to find the value of 100-year $H_S$ relatively difficult to estimate from a small sample of data. Moreover, we expect variability in 100-year $H_S$ (a) between different locations in the same neighbourhood, (b) between locations in different regions, (c) between different climate forcing scenarios, and (d) between different GCMs. In particular, the inter-regional comparison is expected to provide insight into the effect of tropical cyclones (TCs) on extremal predictions from projected model data, as extreme sea states EoM are expected to be dominated by TCs, whereas those SoA are not (see Section~\ref{Sct:Dat:CMIP5} for further discussion). The long piControl output allows us to assess the inherent variability of estimates of 100-year $H_S$ carefully, based on EV modelling of data corresponding to a stationary climate using of different lengths. piControl data therefore also provide an essential basis for assessing the material significance of changes found in the projections under different forcing scenarios. We emphasise that only once the inherent steady-state variability of the extreme ocean environment has been quantified, can changes in that extreme environment due to climate forcing be judged fairly. We devote the whole of Section~\ref{Sct:StdStt} to assessing the inherent variability in estimates of 100-year $H_S$ in the piControl data.

Return values are estimated using EV analysis of both peaks over threshold (POT) and block maxima (BM) data isolated from the GCM output. The quantification of trends in extreme $H_S$ with time is achieved by estimating simple non-stationary EV models, the parameters of which vary linearly in time, using Bayesian inference as described in Section~\ref{Sct:Eva}. Since we anticipate that the effects of climate change will be small relative to the uncertainty in estimated model parameters due to small sample size, the assumption of a linear time trend is appropriate. Moreover, estimating a non-stationary EVs model over a long period of time is statistically more efficient that estimating a series of stationary models over shorter time slices of data, the approach favoured by \cite{MccEA20}. Specifically, \cite{Vnm15} found that statistically significant changes in extreme waves can be identified using a non-stationary analysis of historical and future periods together, but not when stationary models are fitted to each period separately and compared. We therefore expect that our non-stationary models will be advantageous relative to the approach of \cite{MccEA20} in examining climate change effects in CMIP5 output in particular. We also anticipate that our approach of estimating different non-stationary EV models independently per GCM using all relevant data, and then aggregating estimates for different GCMs to form a final view, is more appropriate that the approach of mixing partly-standardised data from different GCMs favoured by \cite{MccEA20}. 

\subsection{Objective and outline of article}
The objective of this article is to provide the met-ocean practitioner with a careful assessment of the strength of evidence in support of climate change effects on $H_S$ from CMIP5- and CMIP6-derived wave output data. The layout of the article is as follows. Section~\ref{Sct:Dat} introduces the GCM data, and  Section~\ref{Sct:Eva} outlines the non-stationary EV methods used. We focus on analysis of POT for reasons of increased statistical efficiency, but also consider inferences based on analysis of BM. Section~\ref{Sct:Eda} provides a few illustrative plots of the GCM data. It should be noted that the quantity of data available for analysis (across different GCMs, forcing scenarios, locations and EV modelling) is huge, and we cannot hope to illustrate the data and the analysis adequately in one article. For this reason, the authors have prepared a supplementary material (SM) document (\citealt{EwnJnt22}) to accompany the main text here. The SM document provides a large number of figures and descriptions, referenced throughout this article using the prefix ``SM''; thus, supporting figures for Section~\ref{Sct:Eda} are provided in SM4. As emphasised above, a quantitative assessment of the inherent variability of GCM output corresponding to no climate forcing is an important basis to frame out assessment of climate change under different forcing scenarios. This analysis is reported in Section~\ref{Sct:StdStt} for CMIP6 piControl output corresponding to a time period of 700 years, incorporating EV analysis using models assumed either stationary or non-stationary in time. Section~\ref{Sct:TmpTrn} then estimates the strength of evidence for changes in return value in time for the GCM output, primarily under RCP4.5 and RCP8.5 climate scenarios, but also for the extended set of CMIP6 scenarios. Supporting descriptions for the statistical modelling in Section~\ref{Sct:Eva} are provided in the appendices. Note that for brevity, the term ``GCM'' is used generically to refer to the source of wave output for all CMIP5 and CMIP6 models (whereas of course the wave output for CMIP5 is from a WAVEWATCH-III wind-wave model, forced by winds from a GCM). For the purposes of the current work, we also treat the forcing scenarios ``RCP4.5'' (for CMIP5 models) and ``SSP245'' (CMIP6) are comparable, since they assume the same extent (4.5 Wm$^{-1}$) of radiative forcing achieved in 2100. We treat scenarios ''RCP8.5''  and ``SSP585'' as comparable since they both achieve 8.5 Wm$^{-1}$ forcing by 2100. Hence for brevity, we sometimes refer to the ``RCP4.5'' or ``RCP8.5'' scenario for all GCMs, on the understanding that when applied to a CMIP6 model, we refer to the corresponding SSP scenario.

\FloatBarrier
\section{Data sources} \label{Sct:Dat}
%

\subsection{CMIP5 global coupled models} \label{Sct:Dat:CMIP5}
The CMIP5 data considered in this study are drawn from those used by \cite{MccEA20}. They consist of WAVEWATCH-III model output, with wind forcing from seven GCMs: ACCESS1.0, BCC-CSM1.1, GFDL-CM3, HadGEM2-ES, INMCM4, MIROC5, and MRI-CGCM3. Wave output including $H_S$ are produced at a spatial resolution of 1\textsuperscript{o} degrees, at 6-hourly intervals (\citealt{HmrEA12}). Data include a historical 27-year period (1979–2005), a mid-21\textsuperscript{st} century 20-year period (2026-2045) projection, and an end-21\textsuperscript{st} century 20-year period (2081-2100) projection. Output for two different RCPs are provided: RCP4.5, intermediate emissions scenario, and RCP8.5, high-emissions scenario, for each projection period. \cite{MccEA20} found that the global distribution of mean and 99\textsuperscript{th} percentile $H_S$ values from these seven models compared favourably against those from the much larger analysis of output from 83 models by \cite{MrmEA19}, providing confidence that the subset of seven models is representative of the larger ensemble. 

These is general concern regarding the extent to which the GCMs considered adequate characterise tropical cyclone (TC) events. \cite{MccEA20} report that the BCC-CSM1.1, MIROC5, and MRI-CGCM3 models capture at least 80\% of TCs, as well as winds with speeds more than 30 m/s, when compared with observations made by \cite{ShmEA17} for the western North Pacific region. But in general, the seven-model ensemble was found to only have limited accuracy in representing TC events and tended to underestimate extremes in TC regions. Therefore, whereas some information on TCs can be inferred from the 7-member output considered here, care should be taken in interpreting changes to estimated return values for $H_S$ in TC regions.

In this work, we choose to focus on two interesting geographic regions for detailed analysis, the selection of which is informed by Figure 2 of \cite{MccEA20}, a global map showing changes in 100-year $H_S$ between historical and end-21\textsuperscript{st} century periods. We are particularly interested in considering regions within which a strong climate change effect on 100-year $H_S$ might be anticipated. Accordingly, we choose data for a region EoM where \cite{MccEA20} report a significant decrease of more that 10\% in 100-year $H_S$, and a region SoA showing a significant increase of more than 10\% in 100-year $H_S$.

\begin{figure}
	\centering
	\includegraphics[scale=0.80]{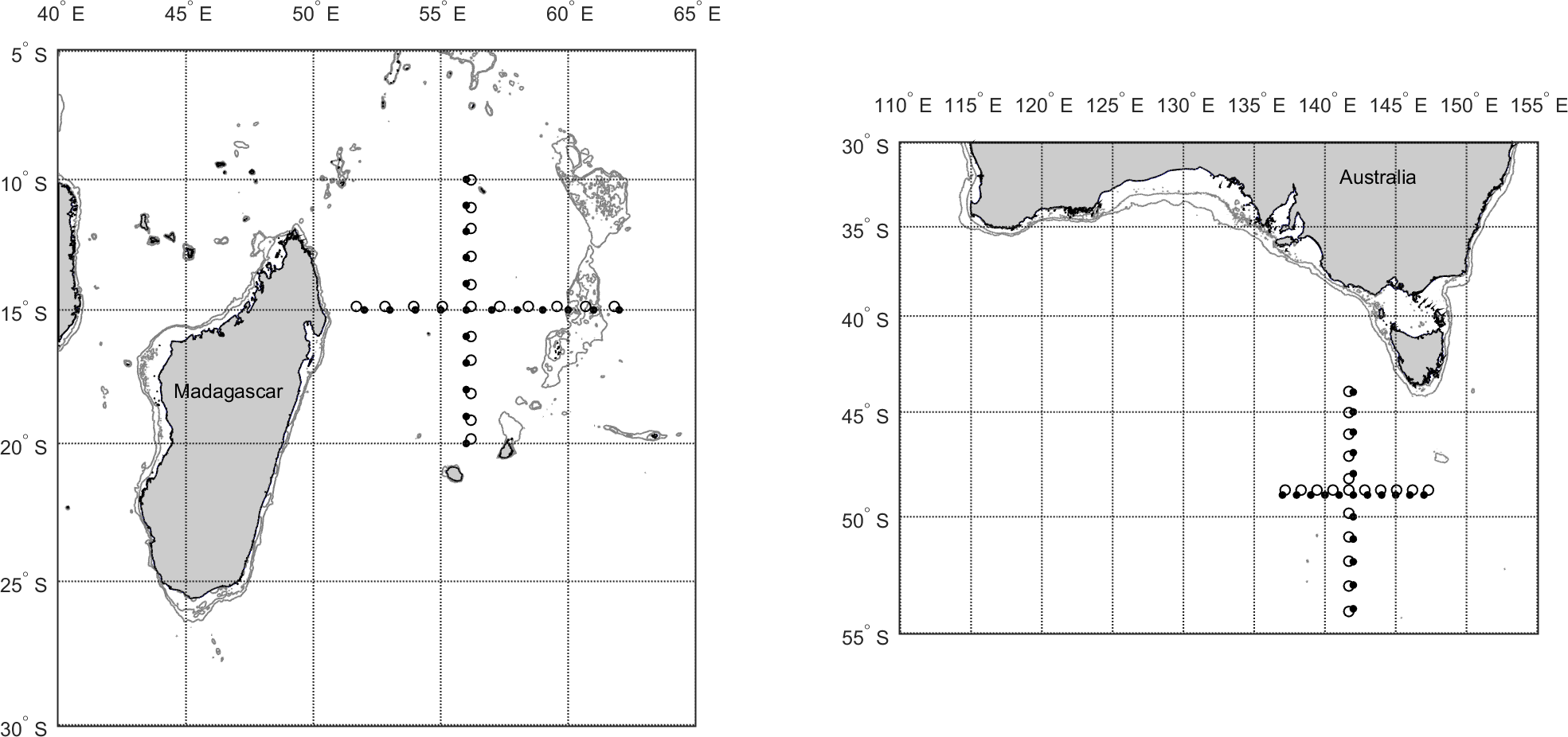}
	\caption{Locations of EoM (left) and SoA (right) grid points used. Dots indicate CMIP5, and circles CMIP6. Grey lines are bathymetric contours at 100m and 1,000m water depth.}
	\label{Fgr:Map}
\end{figure}

In each region we isolate $H_S$ data for meridional and zonal transects of grid points with a common centre location, as shown in Figure~\ref{Fgr:Map}. 
For each GCM, we then perform a non-stationary EV analysis independently per grid point and per GCM. This procedure allows us to assess spatial trends in time changes of extreme $H_S$ per GCM, as well as compare different GCMs. Grey lines in Figure~\ref{Fgr:Map} correspond to contours of water depth at 100m and 1,000m. All locations SoA correspond to water depths greater than 1,000 metres. This is generally true EoM, although some of the more eastward locations along the zonal transect are in the more shallow water of the Nazareth Bank. Figure SM2.1 (in Supplementary Material, \citealt{EwnJnt22}) illustrates time-series of peaks over threshold (POT, top row) and annual maxima (AM, bottom row) of $H_S$ from the ACCESS CMIP5 GCM for forcing scenario RCP4.5, at the centre location of EoM (left) and SoA (right) neighbourhoods.

\subsection{CMIP6 global coupled models} \label{Sct:Dat:CMIP6}
The CMIP6 data considered are taken from output of the FIO‐ESM v2.0 model (\citealt{BaoEA20}, \citealt{SngEA20}), and consist of $H_S$ values for grid points nearest to those of the selected CMIP5 transects, shown in Figure~\ref{Fgr:Map}. Seven CMIP6 experiments are considered (see Table 2 of \citealt{SngEA20}) : a 700-year pre-industrial period (piControl: nominal years 301-1000), a 165-year historical period (1850-2014), three 86-year future scenarios (SSP125: RCP 2.5, SSP245: RCP 4.5, and SSP585: RCP 8.5, all for the years 2015-2100), and two 150-year climate sensitive experiments (1pctCO2: 1\% increase in atmospheric CO$_2$ per year, abrupt-4xCO2: 4\% immediate and abrupt increase of CO$_2$ from the piControl value, both for nominal years 301-450). The $H_S$ values from the FIO‐ESM v2.0 model were available at 3-hourly time steps over a grid varying from 0.27\textsuperscript{o} latitude at the equator to 0.54\textsuperscript{o} latitude in the far north, and 1.125\textsuperscript{o} longitude.

The spatial resolution for EoM varies between 0.27\textsuperscript{o} to 0.36\textsuperscript{o} in degrees latitude, and 1.25\textsuperscript{o} degrees longitude. Hence, the latitude resolution is in the order of the 0.25\textsuperscript{o} suggested by \cite{TmmEA17} to be needed to resolve TCs, but the longitude resolution is not. The spatial resolution for SoA is 0.53\textsuperscript{o} in degrees latitude and 1.25\textsuperscript{o} degrees longitude. Figure SM2.2 illustrates the corresponding data to Figure SM2.1, for the CMIP6 model output.

\subsection{Data preprocessing} \label{Sct:Dat:Prp}
For POT analysis, storm events are identified by isolating continuous intervals of $H_S$ between consecutive up- and down-crossings of a pre-specified threshold level. The threshold level is specified so that more than 20 but not exceeding 25 peaks are selected per annum, for each data set. For analysis of BM, events (corresponding to blocks of length e.g. one or five years, i.e AM or 5-year maxima 5YM respectively) are isolated in the obvious way.

\FloatBarrier
\section{Extreme value analysis} \label{Sct:Eva}
%
We consider two non-stationary EV models in the current work. The first uses generalised Pareto (GP) regression to estimate models for observations of POT, and the second uses generalised extreme value (GEV) regression to estimate models for observations of BM for time blocks of e.g. one year.  For both GP and GEV models, we assume that all model parameters vary linearly over the period of observation, unless stated otherwise. That is, for any model parameter $\eta$, we assume that
\begin{eqnarray}
	\eta_t = \eta(t) = \eta^S + \frac{t}{P} (\eta^E -\eta^S), \text{   for   } t \in (0,P] \label{E:CvrTim}
\end{eqnarray}
in year $t$, where $\eta^S$ and $\eta^E$ are the parameter values at the start year (e.g. 1979) and end year (e.g. 2100) of $P$ years of data to be estimated. 

\subsection{Non-stationary peaks over threshold} \label{Sct:Eva:Gp}
For the GP model, we assume access to observations $\{x_{t_i},t_i\}_{i=1}^{n}$ of peaks over threshold (POT) events $X_t$ at times $t_i \in (0,P)$. We assume that $X_t|X_t>\psi_t$ follows the GP distribution with threshold parameter $\psi_t \in \mathbb{R}$, scale $\sigma_t>0$ and shape $\xi_t \in \mathbb{R}$ for $t \in (0,P)$ with distribution function
\begin{eqnarray}
	F_{\text{GP}}(x|X_t>\psi_t, \psi_t, \sigma_t, \xi_t) = 1 - \left[ 1 + \frac{\xi_t}{\sigma_t}\left(x-\psi_t\right)\right]^{-1/\xi_t}
	\label{E:GpCdf}
\end{eqnarray}
when $\xi_t \neq 0$ and $1-\exp(-(x-\mu_t)/\sigma_t)$ otherwise. Model parameters $\eta_t \in \{\sigma_t, \xi_t\}$ vary with $t$ as described in Equation~\ref{E:CvrTim}. To use Equation~\ref{E:GpCdf} in practice also requires a model for EV threshold $\psi_t$. Here we use quantile regression with lack-of-fit criterion
\begin{eqnarray}
	\ell_{\psi}=\tau \sum_{i, r_{i} \geq 0}^{n}\left|r_{i}\right|+(1-\tau) \sum_{i, r_{i}<0}^{n}\left|r_{i}\right| 
	\label{E:GpQnt}
\end{eqnarray}
for residuals $r_i=x_{t_i}-\psi_{t_i}$, and some fixed quantile non-exceedance probability $\tau \in [0,1]$. Equation~\ref{E:GpQnt} can be interpreted as a Laplace likelihood for estimation. Since estimation of an optimal $\tau$ is problematic in general, models are estimated for a wide range of values of $\tau$ exceeding the mode of the empirical distribution of $X_t$, and sensitivities of inferences to $\tau$ assessed. In this work, we examine the performance of GP models over four choices of EV threshold. These are specified in terms of the non-exceedance probability (NEP) to which they correspond, and referred to henceforth as NEP1-4. NEP1 corresponds to $\tau=0.5$, and NEP4 to the non-exceedance probability which leaves 30 threshold exceedances remaining for EV analysis. Values of $\tau$ for intermediate NEP2 and NEP3 are then equally spaced (on log scale) between those of NEP1 and NEP4. The asymptotic GP model form is likely to be more appropriate for NEP4 in general, and hence the bias of inference smallest; but the uncertainty in parameter estimates is likely to be lowest for NEP1. A classic bias-variance trade-off. To use Equation~\ref{E:GpCdf} for return value estimation, we also need to estimate the annual rate of occurrence $\rho_t$ of threshold exceedances in time for given $\tau$. We achieve this using Poisson regression (e.g. \citealt{ChvDvs05}, \citealt{RssEA17}) with density 
\begin{eqnarray}
	f(\{c_t\} \mid \rho_t)=\exp \left(-\sum_{t=1}^{P} \rho_{t}\right) \prod_{t=1}^{P} \rho_{t}^{c_{t}}
\end{eqnarray}
where $\{c_t\}_{t=1}^P$ are empirical annual counts of threshold exceedances, and $\rho_t$ is also parameterised as in Equation~\ref{E:CvrTim}. 

For GP-distributed threshold exceedances, and Poisson-distributed rate of threshold exceedance, the distribution of the annual maximum is known to be GEV-distributed (e.g. \citealt{JntEwn13}). Hence, in the absence of parameter uncertainty, the $T$-year return value $Q_t$ at year $t$ is estimated as the $p=1-1/T$ quantile of this distribution. Specifically
\begin{eqnarray}
	Q_t =  \frac{{\sigma}_t}{{\xi}_t} \left[ \left(-\frac{\log p}{\rho_t}\right)^{-{\xi}_t} -1\right] + {\mu}_t   \label{E:GpRV}
\end{eqnarray}
when $\xi_t \neq 0$ and $\mu_t-\sigma_t\log[-(1/\rho_t) \log p ]$  otherwise. Note that since all model parameters change in time, then so does the value of $Q_t$.

Parameter estimation is performed sequentially using Bayesian inference. First we perform quantile regression, generating a sample $\{\hat{\psi}^S_k,\hat{\psi}^E_k\}_{k=1}^{n_I}$ of size $n_I$ from the joint posterior distribution of EV threshold parameters. We then use the non-stationary threshold corresponding to posterior mean parameter estimates from the quantile regression, (a) in Poisson regression, to generate a sample $\{\hat{\rho}^S_k,\hat{\rho}^E_k\}_{k=1}^{n_I}$ from the joint posterior of $\rho$; and (b) in GP regression to sample $\{\hat{\sigma}^S_k,\hat{\sigma}^E_k,\hat{\xi}^S_k,\hat{\xi}^E_k\}_{k=1}^{n_I}$ from the joint posterior of GP parameters, where $n_I>10000$. These sets of posterior parameters are used to estimate the distribution of $T$-year return value, and in particular to compare the estimates $Q_1$ (for the first year) with $Q_P$ (for the last year) with $T=100$. \ref{Sct:App:GpUnc} provides a discussion on the merits of adopting the posterior mean of quantile regression parameters as the basis for subsequent inference. Figures SM3.1-2 illustrate typical non-stationary EV fits to CMIP5 and CMIP6 samples of POT. Figure SM3.3 gives posterior densities of 100-year POT $H_S$ for the start year (1979) and end year (2100) from an EV model for CMIP5 POT output.

\subsection{Non-stationary block maxima} \label{Sct:Eva:Gev}
For EV estimation of block maxima (BM), we suppose we have access to observations $\{x_{t_i},t_i\}_{i=1}^{\lfloor P/b \rfloor}$ of BM events $X_{b,t}$ for blocks of $b$ years in length. For annual maxima ($b=1$) the corresponding observation times $t_i$ are then $t_i=i$, $1,2,...,P$. We assume that $X_{b,t}$ follows the GEV distribution with non-stationary location parameter $\mu_t \in \mathbb{R}$, scale $\sigma_t>0$ and shape $\xi_t \in \mathbb{R}$ for $t \in \{1,2,...,P\}$ with distribution function
\begin{eqnarray}
	F_{GEV}(x|\mu_t, \sigma_t, \xi_t) = \exp\left\{ - \left[ 1 + \frac{\xi_t}{\sigma_t}\left(x-\mu_t\right)\right]^{-1/\xi_t} \right\} \label{E:GevCdf}
\end{eqnarray}
when $\xi_t \neq 0$ and $\exp(\exp(-(x-\mu_t)/\sigma_t))$ otherwise. Model parameters $\eta_t \in \{\mu_t, \sigma_t, \xi_t\}$ vary with $t$ are described in Equation~\ref{E:CvrTim}. The $T$-year return value $Q_t$ for year $t$ is estimated as the $p=1-b/T$ quantile of $F_{GEV}(x;t)$, so that
\begin{eqnarray}
	Q_t =  \frac{{\sigma}_t}{{\xi}_t} \left[ \left(-\log p\right)^{-{\xi}_t} -1\right] + {\mu}_t  \label{E:GevRV}
\end{eqnarray}
when $\xi_t \neq 0$ and $\mu_t-\sigma_t \log(-\log p)$ otherwise. Parameter estimation is performed using Bayesian inference as described in \ref{Sct:App:BysInf}, yielding a sample of $n_I$ joint posterior estimates $\{\hat{\mu}^S_k,\hat{\mu}^E_k,\hat{\sigma}^S_k,\hat{\sigma}^E_k,\hat{\xi}^S_k,\hat{\xi}^E_k\}_{k=1}^{n_I}$ where $n_I>10,000$. These can be used to estimate the empirical distribution of quantities of interest, such as $Q_1$ and $Q_P$, and compare them.

\subsection{Stationary analysis} \label{Sct:Eva:Stt}
For the stationary EV analysis considered in Section~\ref{Sct:StdStt}, all model GP and GEV parameters $\eta$ are assumed not to vary with time. Note that if $X$ (Section~\ref{Sct:Eva:Gp}) and $X_b$ (Section~\ref{Sct:Eva:Gev}) correspond to peaks over threshold and BM of the same physical process, then we expect relationships between the sets of corresponding GEV and GP parameters. We do not consider these relationships in this work. 

\FloatBarrier
\section{Exploratory data analysis} \label{Sct:Eda}
\begin{figure}[h!]
	\centering
	\includegraphics[width=0.70\columnwidth]{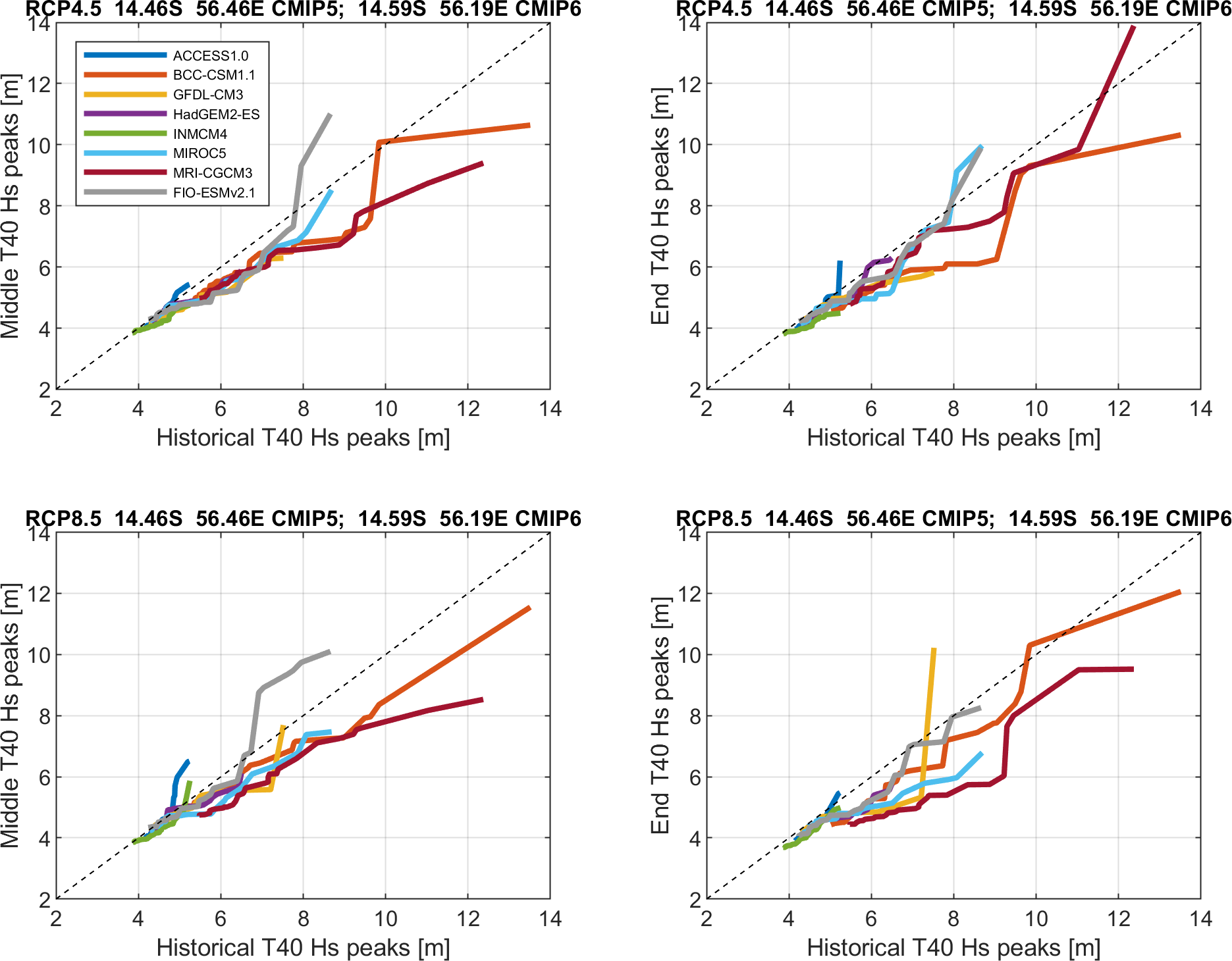}\\
	\caption{Comparison of the ordered 40 largest values (T40) of POT $H_S$ for the ``Historical'' (1979–2005), ``Middle'' (2026-2045) and ``End'' (2081-2100) time periods, for all CMIP5 GCMs, and the FIO-ESM CMIP6 GCM, for forcing scenario RCP4.5 (and SSP245, top row) and RCP8.5 (and SSP585, bottom row) at the centre location EoM. Left-hand panels give scatter plots for the Middle period on the Historical period; right-hand panels give the corresponding plots for the End period on the Historical period. The line $y=x$ is added for guidance.}
	\label{Fgr:CmpPotTalEoM}
\end{figure}

\begin{figure}[h!]
	\centering
	\includegraphics[width=0.70\columnwidth]{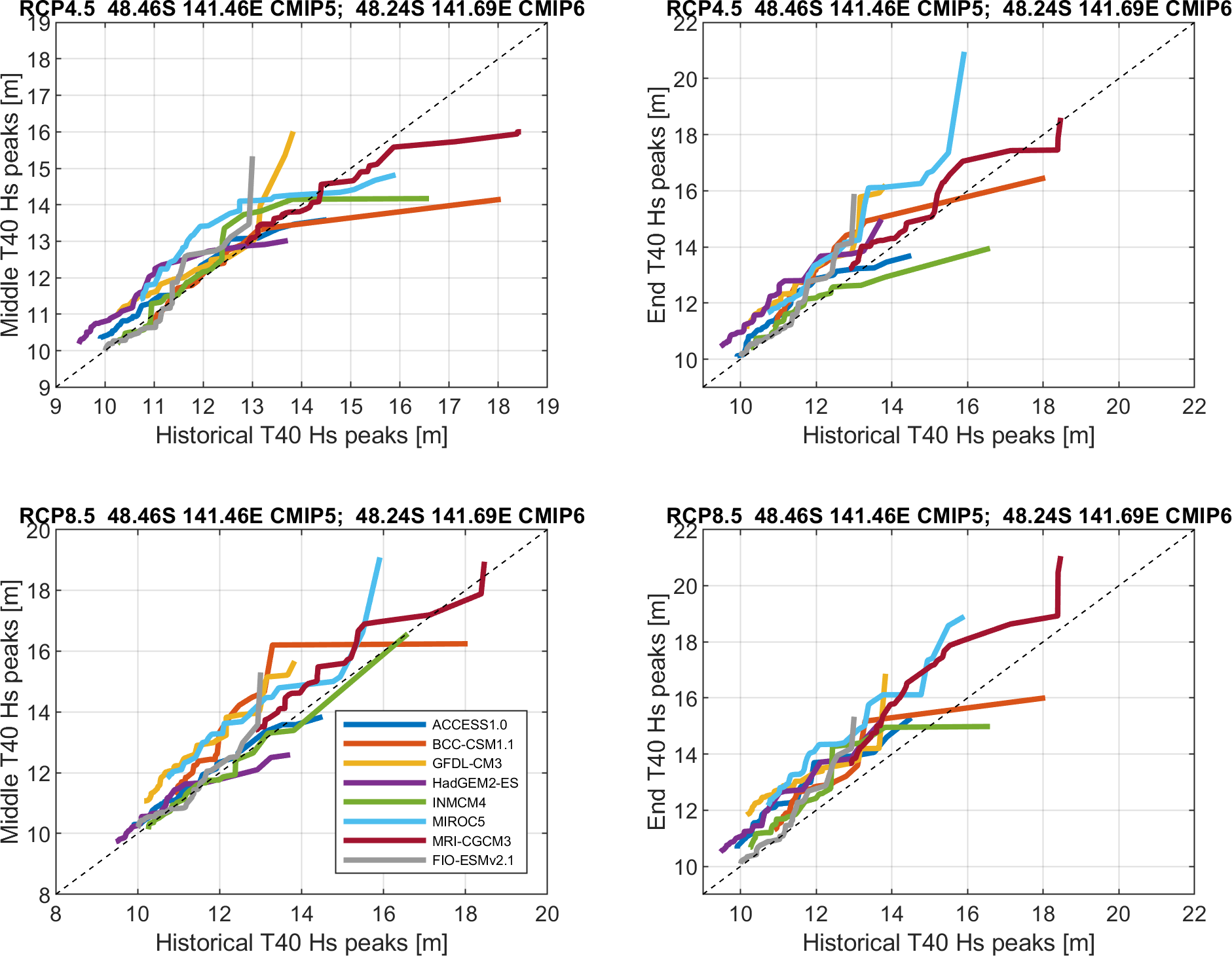}\\
	\caption{Comparison of the ordered 40 largest values of POT $H_S$ for the Historical, Middle and End time periods, for all CMIP5 GCMs, and the FIO-ESM CMIP6 GCM, for forcing scenario RCP4.5 (top row) and RCP8.5 (bottom row) at the centre location SoA. Left-hand panels give scatter plots for the Middle period on the Historical period; right-hand panels give the corresponding plots for the End period on the Historical period.The line $y=x$ is added for guidance.}
	\label{Fgr:CmpPotTalSoA}
\end{figure}

%
In this section, we illustrate some features of the CMIP5 and CMIP6 data considered, referring the reader to supporting exploratory analysis in SM4. Given the time structure of the CMIP5 data partitioned into ``Historical'' (1979–2005), ``Middle'' (2026-2045) and ``End'' (2081-2100) time periods, and our primary interest in changes in extremes of $H_S$, it is interesting to start by comparing the tails of empirical distributions of POT $H_S$ across GCMs. This comparison involves no model fitting, but rather simply plotting sorted values of POT $H_S$ from different times periods against each other. Figures~\ref{Fgr:CmpPotTalEoM} and \ref{Fgr:CmpPotTalSoA} compare the sorted 40 largest values of POT $H_S$ for the Middle and End time periods with the sorted 40 largest values for the Historical period, for all CMIP5 GCMs, and the (same time periods of) FIO-ESM CMIP6 GCM, for forcing scenario RCP4.5 (and SSP245, top row) and RCP8.5 (and SSP585, bottom row) at the centre location EoM (Figure~\ref{Fgr:CmpPotTalEoM}) and SoA (Figure~\ref{Fgr:CmpPotTalSoA}). The left-hand panels of the figures show scatter plots for the Middle period on the Historical period, and the right-hand panels give the corresponding plots for the End period on the Historical period. We can assess the likely effect of climate change on extreme POT $H_S$ simply from the figures. A decrease in POT $H_S$ between periods for a given GCM corresponds to a (coloured) line lying below the y=x dashed line in the panels of the figures. For EoM, we see that this is generally the case, whereas the opposite is generally true for SoA, indicating an increase in extreme POT $H_S$ there. Of course, there is considerably variability between GCMs, and it is apparent that differences between GCMs are at least as large as the effect of RCP choice, or choice of pair of time periods to compare. Nevertheless, there is some evidence from these figures that a systematic trend in extreme POT $H_S$ over time is discernable.
Figures SM4.1-2 in the supplementary material illustrate the analogous analysis using annual maxima (AM) of $H_S$, and in general the features observed are consistent with those described here for POT. Figures SM4.3-34 provide more extensive comparisons along all transect locations for the CMIP5 ACCESS and CMIP6 GCMs; corresponding figures for other CMIP5 GCMs are available from the authors on request. Figures~\ref{Fgr:CmpPotTalEoM}-\ref{Fgr:CmpPotTalSoA} provide a reasonable summary of the trends observed.

\begin{figure}[h!]
	\centering
	\includegraphics[width=0.65\columnwidth]{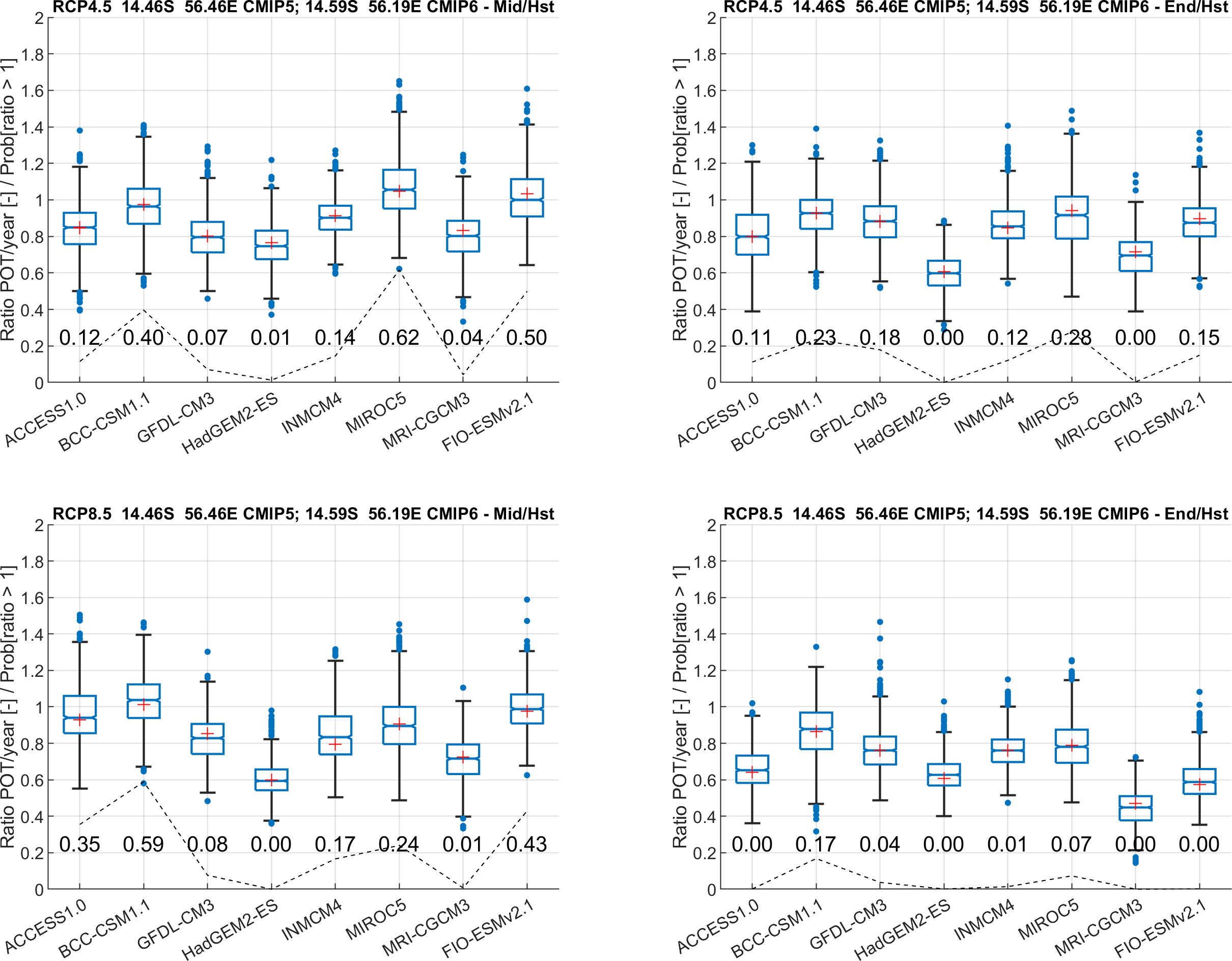}\\
	\caption{Box-whisker plots for ratio of annual rates of POT $H_S$ at centre location EoM, for all CMIP5 and CMIP6 GCMs. Panels show ratios for the the Middle relative to Historical periods (left) and the End relative to Historical periods (right), for RCP4.5 (and SSP245, top) and RCP8.5 (and SSP585, bottom) forcing scenarios, using a POT threshold corresponding to the 80\%ile of the data for the historical period. Each box-whisker indicates the mean (red crosses), median, 25\%, 75\% quartiles (blue lines), the smallest and largest values not exceeding 1.25 $\times$ inter-quartile range from the median (black), and outliers (blue dots). Also shown (dashed black line, and values per GCM) is the estimated probability that the ratio of annual rate exceeds unity.}
	\label{Fgr:CmpPotRatEoM}	
\end{figure}

\begin{figure}[h!]
	\centering
	\includegraphics[width=0.65\columnwidth]{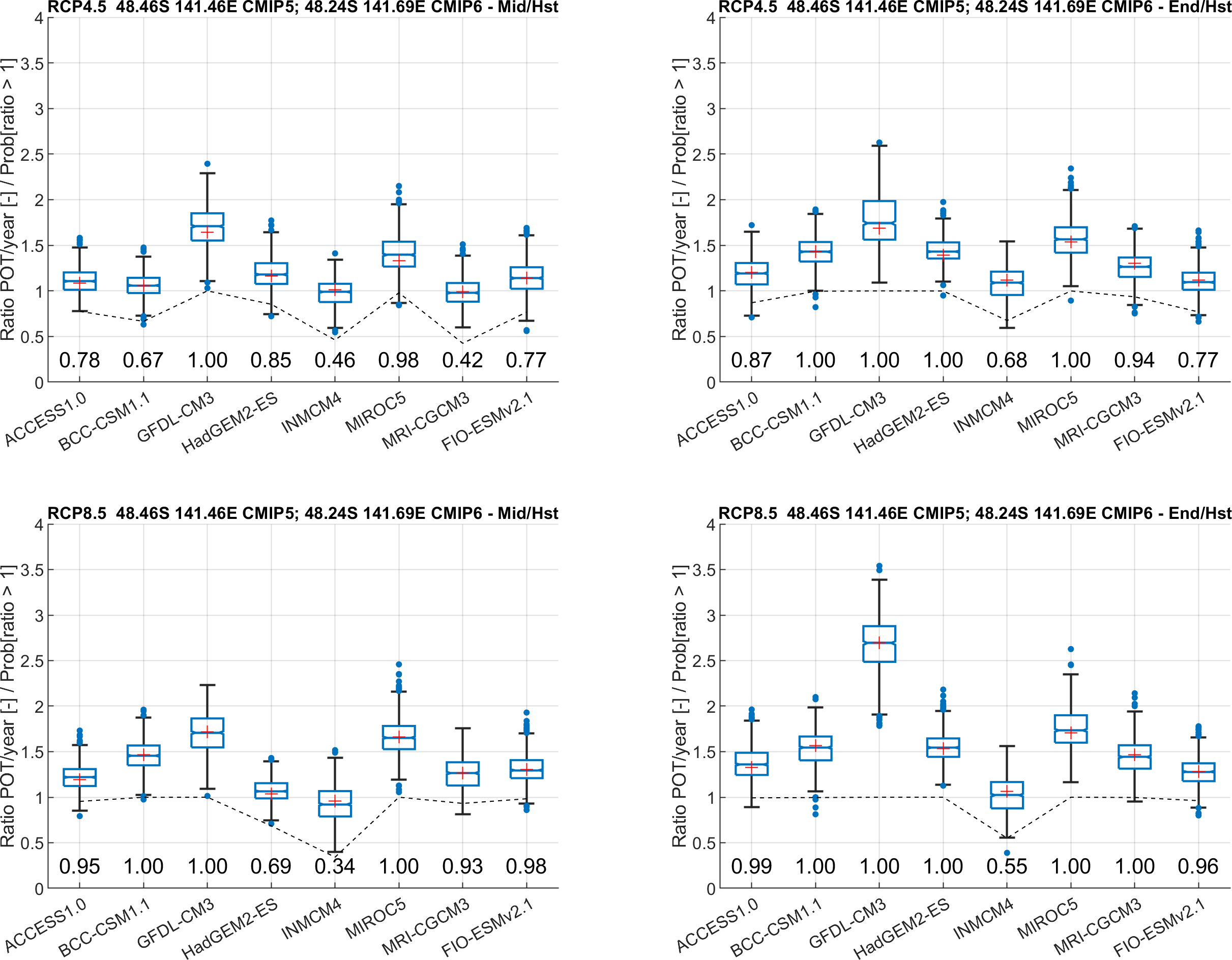}\\
	\caption{Box-whisker plots for ratio of annual rates of POT $H_S$ at centre location SoA, for all CMIP5 and CMIP6 GCMs. Panels show ratios for the the Middle relative to Historical periods (left) and the End relative to Historical periods (right), for RCP4.5 (top) and RCP8.5 (bottom) forcing scenarios, using a POT threshold corresponding to the 80\%ile of the data for the historical period. Each box-whisker indicates the mean (red crosses), the median, 25\%, 75\% quartiles (blue lines), the smallest and largest values not exceeding 1.25 $\times$ inter-quartile range from the median (black), and outliers (blue dots). Also shown (dashed black line, and values per GCM) is the estimated probability that the ratio of annual rate exceeds unity.}
	\label{Fgr:CmpPotRatSoA}
\end{figure}

We can also examine whether the rate at which a particular value of POT $H_S$ is exceeded changes between the different time periods, for different GCMs. This analysis is illustrated in Figures~\ref{Fgr:CmpPotRatEoM} and \ref{Fgr:CmpPotRatSoA} for EoM and SoA. Panels in each plot show ratios of annual rates of exceedance of the level corresponding to the 80\%ile for the historical period, as box-whisker structures (the details of which are explained in the figure captions). In each figure, panels correspond to RCP4.5 (top row) and RCP8.5 (bottom row), and comparison of Middle with Historical (left column) and End with Historical. A bootstrapping scheme is employed to estimate the uncertainties in the rate of POT the figures, implemented as follows: first (a) an EV threshold for the analysis is set equal to the 80\%ile of the distribution of POT $H_S$ over all years of data for the Historical period, then (b) the numbers of exceedances of this level are estimated empirically for each of the Historical, Middle and End periods, and the ratios of rates calculated. Finally (c) we repeat steps (a) and (b) using bootstrap resamples of the full Historical, Middle and End period to estimate uncertainties.

For EoM (Figure~\ref{Fgr:CmpPotRatEoM}) the value of mean and median ratio of rates is around unity or below it, but again there is considerable uncertainty in the ratio of rates within GCM as well as between GCMs. The clearest trend appears for the End:Historical comparison for RCP8.5 forcing scenario. For SoA (Figure~\ref{Fgr:CmpPotRatSoA}) the ratio of rates is generally around unity or larger than it, indicating an increase in the rate of occurrence of large POT $H_S$; again this trend is most pronounced for the End:Historical comparison with RCP8.5 forcing. Each panel of Figures~\ref{Fgr:CmpPotRatEoM} and \ref{Fgr:CmpPotRatSoA} also shows a dashed line and accompanying values, corresponding to the estimated probability that the annual rate of occurrence of extreme POT $H_S$ has increased. In general, for EoM this probability is near zero (although there are notable exceptions); for SoA, the estimated probability is near unity (again with a few exceptions).
Figures SM4.35-39 provide further illustrations, supporting the general trends observed in Figures~\ref{Fgr:CmpPotRatEoM} and \ref{Fgr:CmpPotRatSoA}: the tail of the distribution of $H_S$ becomes shorter with time for EoM, and longer for SoA, under both RCP scenarios.   

\FloatBarrier
\section{Inherent steady state variability} \label{Sct:StdStt}
%
Natural climate variability, including the action of inter-annual and longer-term atmospheric oscillations, results in non-stationary temporal effects in met-ocean databases spanning several decades. Nevertheless, the climate might still be considered stationary in the long term, and it is reasonable to expect that the pre-industrial climate would be stationary (e.g. \citealt{ErnEA15}), representing a period several hundred years prior to the industrial revolution. Thus, pre-industrial data sets offer the possibility to study the effects of natural climate variability on estimates of extremes. In this respect, \cite{BaoEA20} examine the time evolution of net radiative flux parameters from the piControl run and conclude that the FIO‐ESM model reaches the pre-industrial equilibrium state, after a 300-year spin up period. Consequently, the piControl data set might be expected to represent a stationary climate and we use the data set to asses the natural variability of the wave climate at a location. This allows us to assess the inherent variability in wave height extremes when estimated from a typical length of data. In particular, we would like to understand the natural variability, in a stationary climate, of estimates of the 100-year $H_S$ values and their changes, for data lengths corresponding to those typically available for analysis. Our approach to this is in two parts: first to assess the variability associated with the 100-year $H_S$ using a stationary extremal analysis, and then using non-stationary EV models. 

We also estimate how the 100-year $H_S$ might vary due to the more-or-less arbitrary choices made by the practitioner in making that estimate. For example, we examine the serial variation of estimates for 100-year $H_S$ over the 700 years of output, based on EV analysis of a continuous subsample of data corresponding to a specified time interval (henceforth referred to as a ``segment length'' for clarity, and to distinguish this interval from the ``block length'' for analysis of BM). We consider the effect of choice of EV threshold for POT analysis, and the choice of block length for a BM analysis on the resulting estimated return value. In particular, we demonstrate that the $H_S$ data for EoM represents a mixed population of storm types, demanding the use of higher EV thresholds (POT) and block lengths (BM) for reliable estimation of 100-year $H_S$. We also quantify the zonal and meridional variation of return value for EoM and SoA.   

\subsection*{Stationary EV analysis}
\begin{figure}[h!]
	\centering
	\includegraphics[width=0.70\columnwidth]{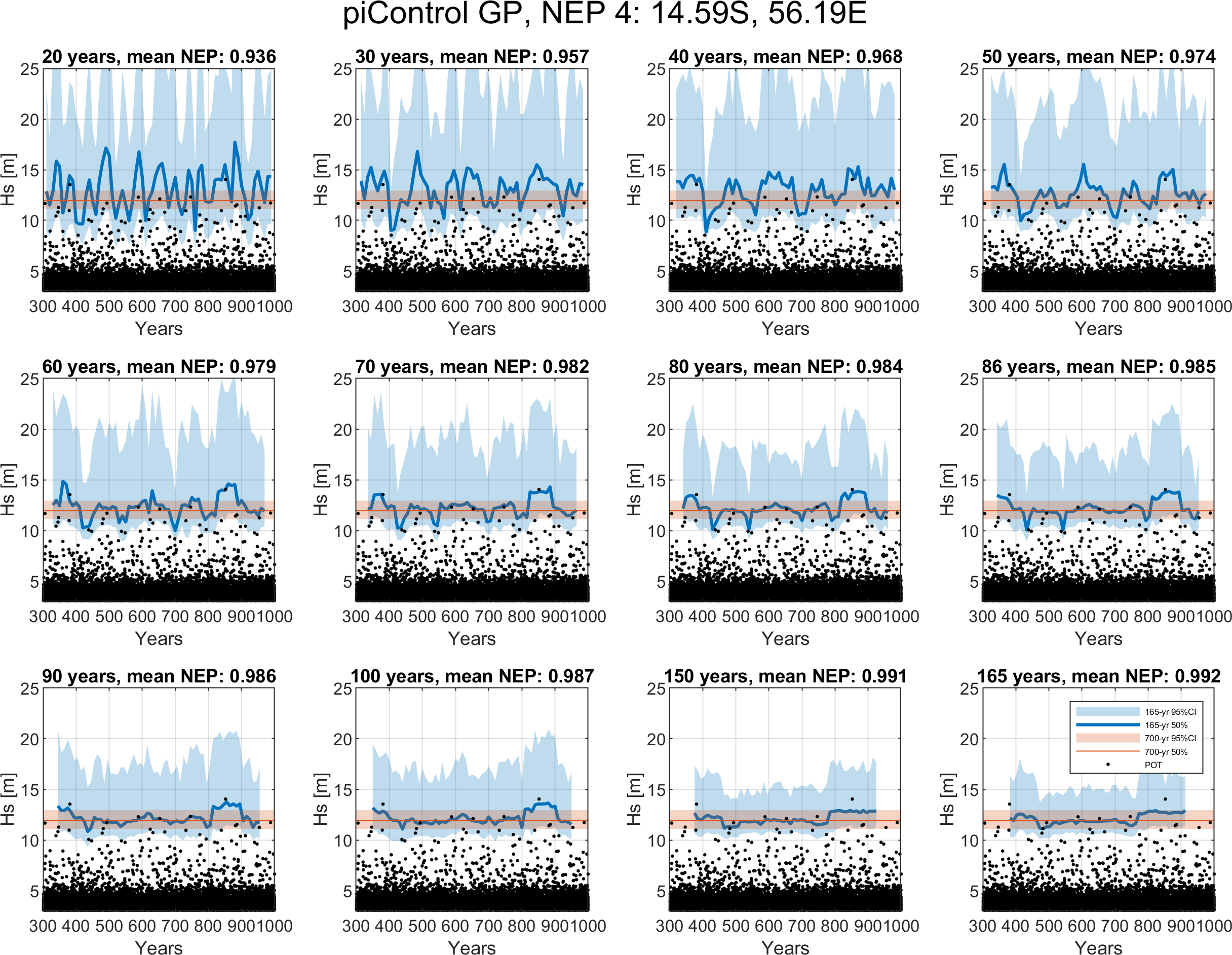}\\
	\caption{Estimates of 100-year $H_S$ in time for the CMIP6 piControl data, using EV threshold NEP4. Each panel gives return value estimates from stationary EV analysis using an interval of data of ``segment length'' specified in panel title, starting at the year indicated on the x-axis, at the centre location EoM. Within each panel, the posterior median estimate is given as a blue line, and its 95\% credible interval (CI) as a blue band. Also shown (in orange, and common to all panels) are the corresponding posterior median and 95\% CI using the full 700 years of data for EV analysis.  In each panel, the sample data is represented as black dots. The panel title also gives the value of non-exceedance probability $\tau$ corresponding to threshold level NEP4. The 95\% credible intervals for shorter segment lengths have been truncated to ease comparison.}
	\label{Fgr:StdStt-SgmLng-EoM}
\end{figure}

The panels of Figure~\ref{Fgr:StdStt-SgmLng-EoM} show the evolution of estimates for 100-year POT $H_S$ based on GP modelling of data corresponding to an interval of time of specific ``segment length'', for the centre location EoM. The posterior distribution of return value from the Bayesian inference is summarised in terms of its median value (blue line) and a central 95\% credible interval (blue band). The corresponding posterior median and central 95\% credible interval, estimated using all 700 years of data, is also shown (orange line and band) in each panel as a benchmark, together with the underlying sample (black dots). In this Figure, EV threshold NEP4 is used admitting only the most extreme values to the analysis. The corresponding figures for NEP1-3 are provided as Figures SM5.1-3. The choice of segment lengths to examine was dictated by the lengths of GCM output available for the various climate scenarios to be considered in Section~\ref{Sct:TmpTrn}: 86 years for the three future scenarios (SSP1-25, SSP245, and SSP585, respectively), 150 years for the two climate sensitive experiments (1pctCO2 and abrupt-4xCO2) simulations, and 165 years for the historical data set. For comparison, we also include return value estimates based on periods increasing in decadal increments from 20 to 100 years corresponding to the lengths of hindcast data sets that might typically be available for estimating Hs100.

\begin{figure}[h!]
	\centering
	\includegraphics[width=0.70\columnwidth]{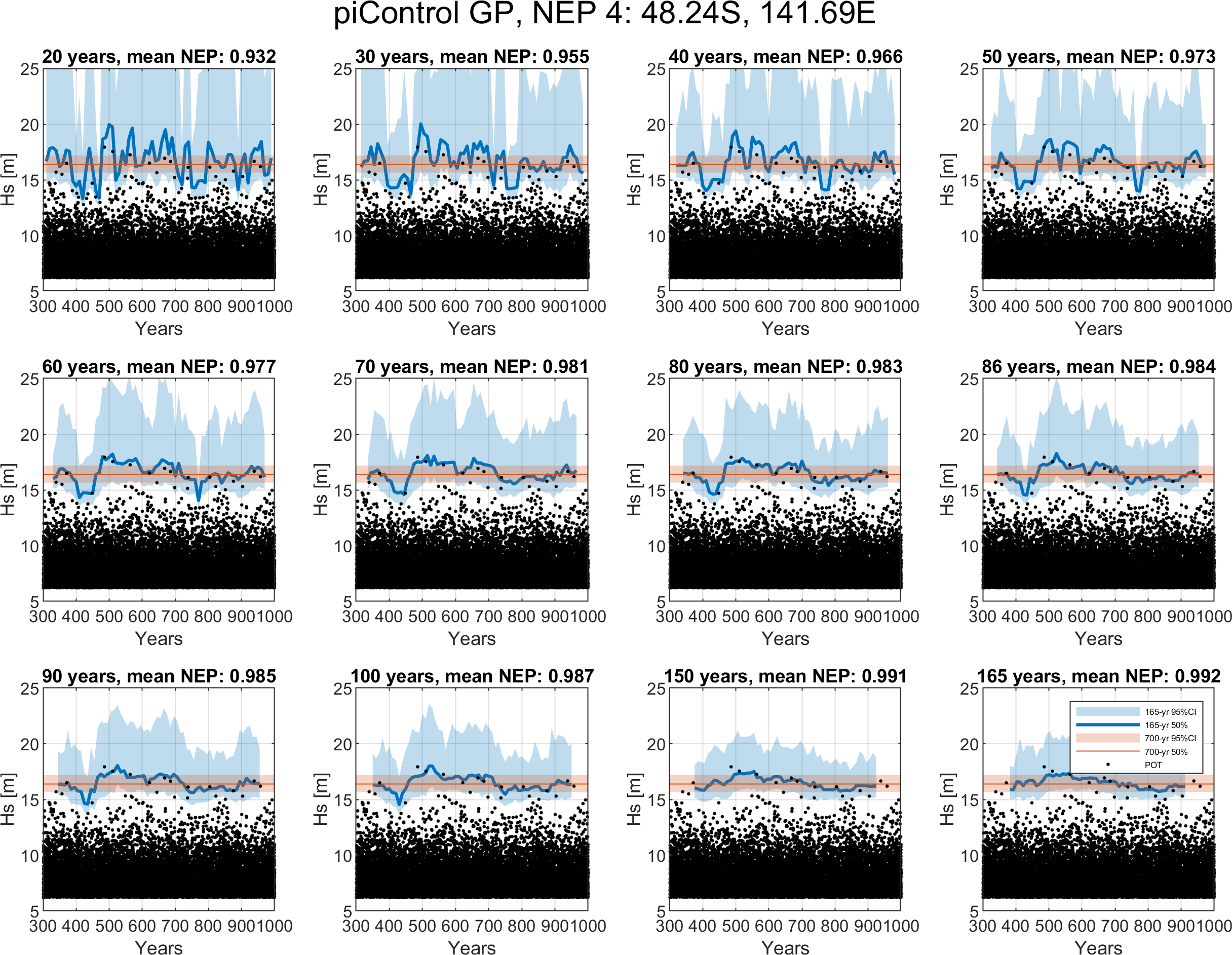}\\
	\caption{Estimates of 100-year $H_S$ in time for the CMIP6 piControl data, using EV threshold NEP4. Each panel gives return value estimates from stationary EV analysis using an interval of data of ``segment length'' specified in panel title, starting at the year indicated on the x-axis, at the centre location SoA. Within each panel, the posterior median estimate is given as a blue line, and its 95\% credible interval (CI) as a blue band. Also shown (in orange, and common to all panels) are the corresponding posterior median and 95\% CI using the full 700 years of data for EV analysis. In each panel, the sample data is represented as black dots. The panel title also gives the value of non-exceedance probability $\tau$ corresponding to threshold level NEP4.}
	\label{Fgr:StdStt-SgmLng-SoA}
\end{figure}

Figure~\ref{Fgr:StdStt-SgmLng-EoM} indicates that the variability in posterior median return value in time is largest for the smallest segment lengths in general; changes in posterior median return value of 5m over a relatively short period of time are present, even for segment lengths as large as 50 years. The 95\% credible interval for return value is long-tailed with positive skew, often exceeding 25m even though the median value is approximately 12m. It appears that the return value estimates for all segment lengths are, taking a long-term temporal view, approximately unbiased relative to the posterior median return value estimated using the full piControl data. The corresponding behaviour for SoA in Figure~\ref{Fgr:StdStt-SgmLng-SoA} (for NEP4, and Figures SM5.5-7 for NEP1-3) is generally similar. Again, for NEP4, estimates of return value are approximately unbiased (over the 700 years) for all segment lengths considered, but the local temporal variability in posterior median return value, and the 95\% credible interval for return value, is again huge for small segment lengths.
Figures SM5.5-7 and Figure~\ref{Fgr:StdStt-SgmLng-SoA} illustrate that increasing the EV threshold has little effect on return value estimate for the centre location SoA, whereas the posterior distribution of return value, and its temporal variability, are smaller for lower non-exceedance probabilities. This suggests that perhaps NEP1 would be an appropriate choice of EV threshold SoA. The same cannot be said from inspection of Figures SM5.1-3 and Figure~\ref{Fgr:StdStt-SgmLng-EoM} for the centre location EoM. Here, for NEP1, NEP3 and NEP4, the posterior median return value is around 12m; but for NEP2, the value is nearer 16m (with wide posterior distributions even for estimation using all 700 years of data). 

Return value estimates using lower non-exceedance probabilities appear to be affected by a mixed population storms at the centre location EoM. To illustrate this effect, the panels of Figure SM5.9 show posterior median tail fits obtained using NEP1-4 there. Only tail fits using NEP3 and NEP4 are reasonable, accurately capturing the convexity of the empirical distribution at around 11m. Fits using NEP1 and NEP2 are poor; the fact that NEP1 provides a reasonable estimate for return value is fortuitous. For comparison, Figures SM5.10 shows the corresponding fits for SoA, all of which appear reasonable, with the possible exception of NEP1. The effect of a mixed population of storms EoM is also observed in the analysis of BM; Figures SM5.11-12 show the posterior median GEV tail fits based on annual maxima (AM), 5-year maxima (5YM for brevity), 10-year maxima and 20 year maxima data. The fit for AM is very poor, but once block length is extended to 5 or more years, an excellent tail fit is obtained. For the SoA region, an excellent tail fit for all block lengths is obtained. We conclude that, if an analysis of BM is desired EoM, block lengths of at least 5 years are necessary to avoid fitting to a mixed population of storms. Physically, we interpret this as the occurrence of tropical cyclone events in the CMIP6 output EoM with a rate of less than one per annum, but generally at least one every five years. Using 5YM data, the effect of segment length on the stationary GEV analysis of piControl output for centre locations EoM and SoA is illustrated in Figure SM5.13. Because we require at least 20 observations to estimate the GEV model, we insist on segment lengths of at least 100 years. The resulting estimates are reassuringly in excellent agreement with those found using POT analysis with NEP4.

\begin{figure}[h!]
	\centering
	\includegraphics[width=0.70\columnwidth]{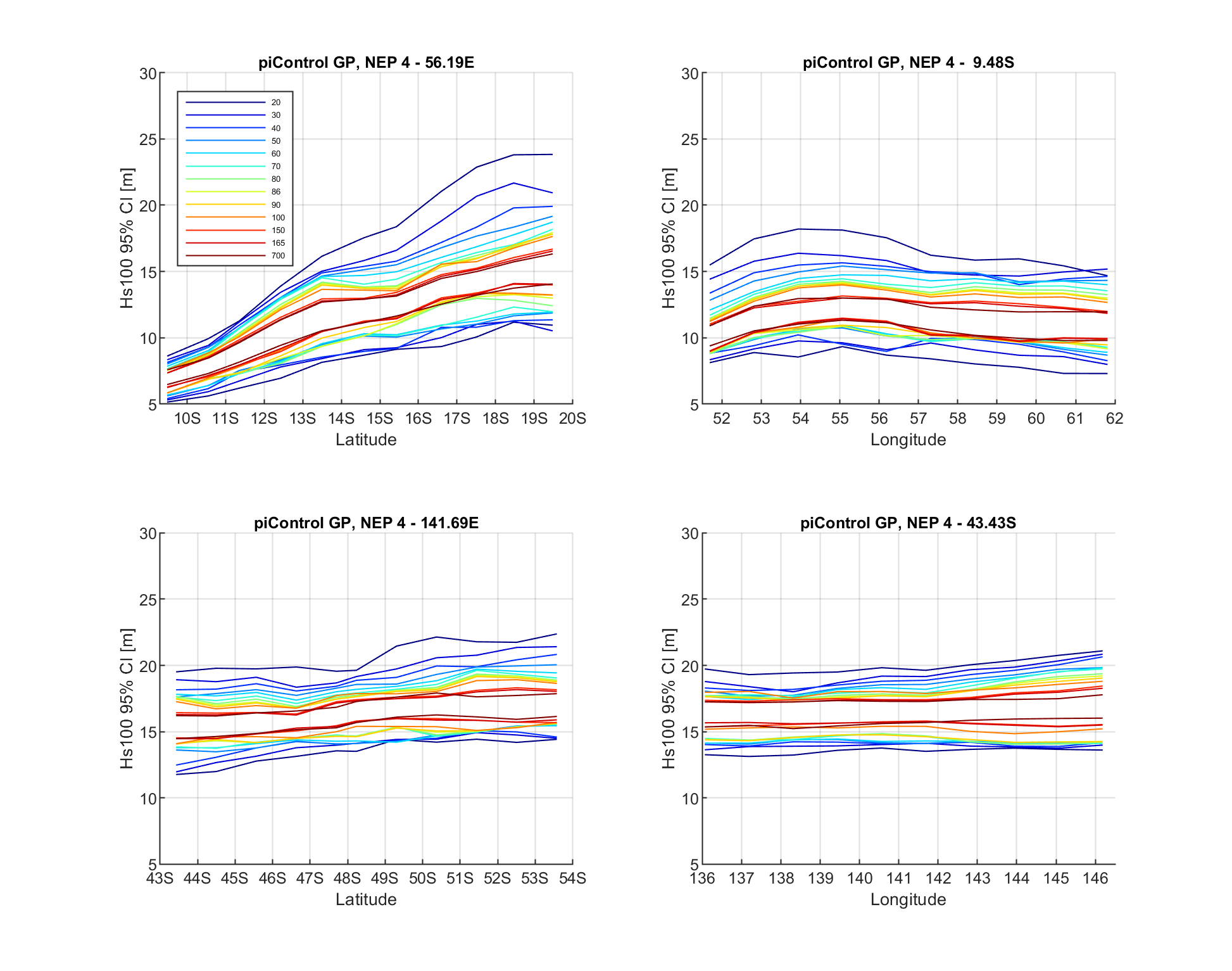}\\
	\caption{Estimates of meridional (left) and zonal (right) variation of 100-year POT $H_S$ for EoM (top) and SoA (bottom) for the CMIP6 piControl data, using EV threshold NEP4. 13 pairs of ordered coloured lines in each panel give 95\% credible intervals for the return value from stationary EV analysis using intervals of data of specific length (in years, as in Figures~\ref{Fgr:StdStt-SgmLng-EoM} and \ref{Fgr:StdStt-SgmLng-SoA}). The calculation procedure is outlined in the text.}
	\label{Fgr:StdStt-LngLtt}
\end{figure}

Figure~\ref{Fgr:StdStt-LngLtt} summarises the meridional and zonal variation of 100-year POT $H_S$ for EoM (top) and SoA (bottom) estimated using stationary GP analysis with threshold level NEP4, in terms of 95\% credible intervals for different choices of segment length from 20 years (dark blue) to 165 years (dark red). The calculation procedure is as follows. For each combination of starting time and segment length, we calculate the posterior median return value from the Bayesian inference. We then calculate the 95\% interval over all steps for the segment length, and refer to this interval as the ``natural 95\% interval'', providing an indicative range for our best estimate of return value for a random start year. There is evidence for increasing return value with distance from the equator as might be expected. There are no strong zonal trends. Supporting plots of meridional and zonal trends for different EV thresholds and segment lengths are given in Figures SM5.14-17. The corresponding zonal and meridional trends using AM and 5YM data are also illustrated in Figures SM5.18-19.

\begin{figure}[h!]
	\centering
	\includegraphics[width=0.70\columnwidth]{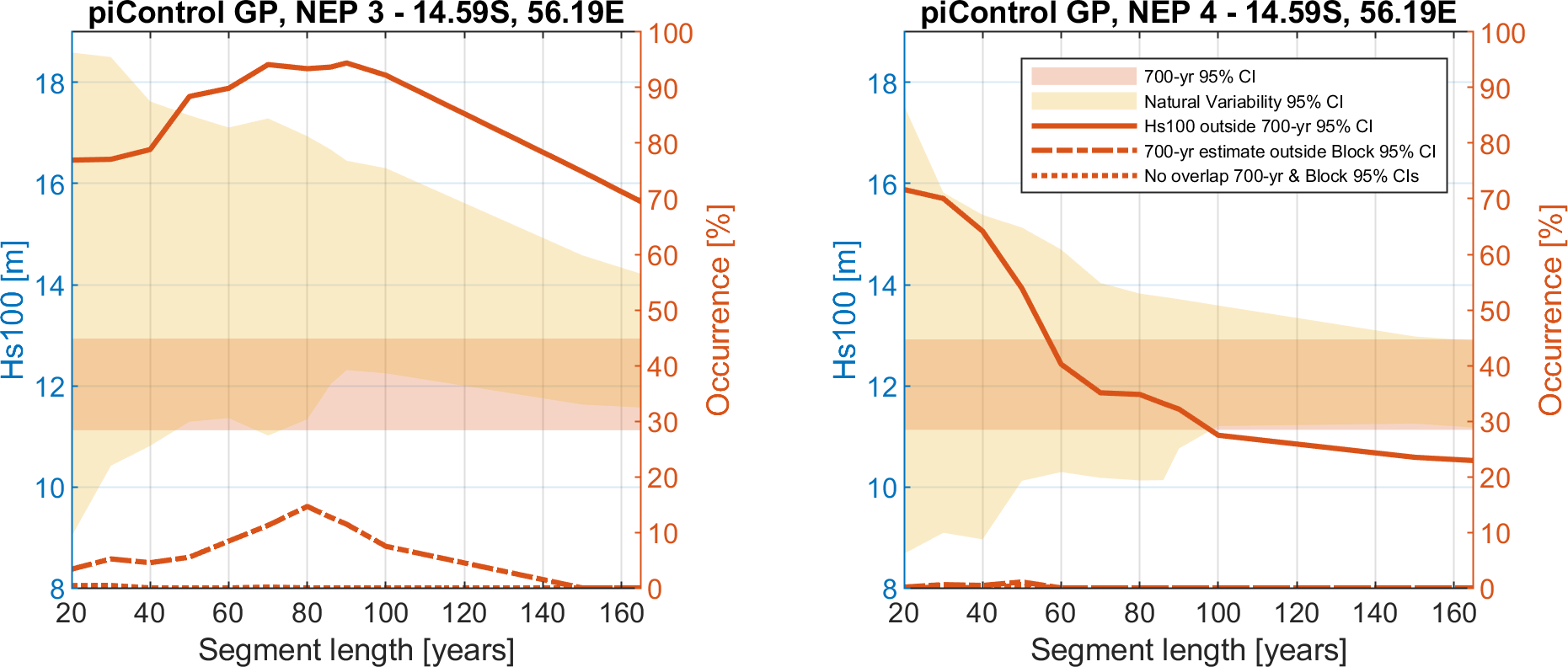}
	\caption{Effect of segment length and threshold on estimate of 100-year return value for POT $H_S$ at centre location EoM. Panels represent estimates for EV threshold levels NEP3-4. In each panel, with respect to the left-hand y-axis, the dark and light orange bands refer respectively to (a) the 95\% credible interval for the return value using the full sample, and (b) the empirical 95\% uncertainty interval of \textit{posterior median} return values for the segment length given on the x-axis. With respect to the right-hand y-axis, solid, dashed and dotted orange lines indicate the percentage of starting years for which (a) the estimated \textit{posterior median} return value for given segment length lies outside the 95\% credible interval estimated using the full sample, (b) the estimated \textit{posterior median} return value using the full sample lies outside the 95\% credible interval estimated using the given segment length, and (c) there is no overlap at all between 95\% credible intervals estimated from the full sample and from a given segment length.}
	\label{Fgr:StdStt-CvrEoM}	
\end{figure}

\begin{figure}[h!]
	\centering
	\includegraphics[width=0.70\columnwidth]{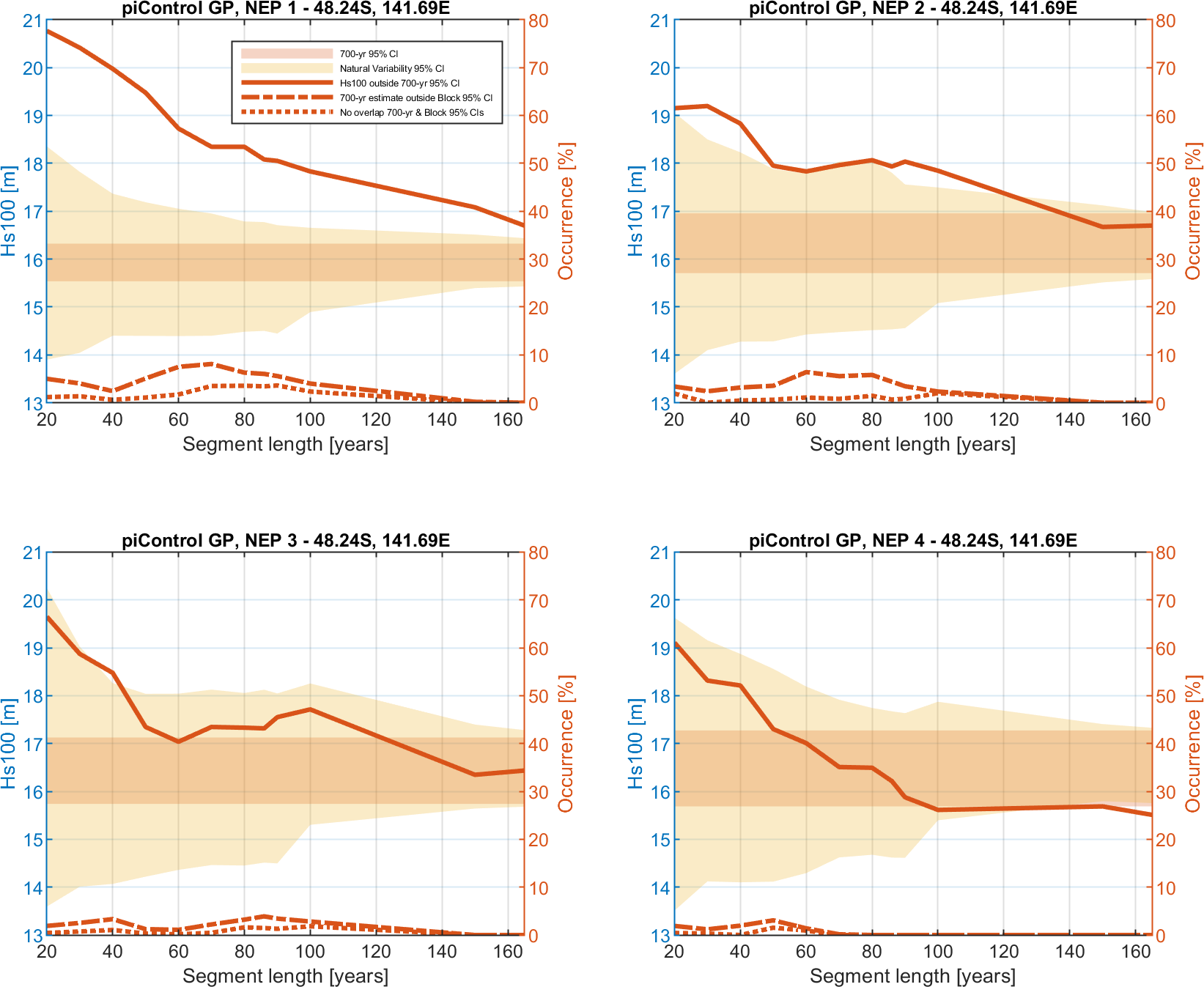}
	\caption{Effect of segment length and threshold on estimate of 100-year return value for POT $H_S$ at centre location South of Australia. Panels represent estimates for EV threshold levels NEP1-4. For details, see Figure~\ref{Fgr:StdStt-CvrEoM}.}
	\label{Fgr:StdStt-CvrSoA}
\end{figure}

Figures~\ref{Fgr:StdStt-CvrEoM} and \ref{Fgr:StdStt-CvrSoA} summarise the uncertainties associated with estimating the 100-year POT $H_S$ from segments with different length and start times from the piControl output, for centre point locations EoM and SoA. In each panel with respect to the left-hand abscissa, dark and light orange bands refer respectively to (a) the 95\% credible interval for the return value using the full sample, and (b) the empirical 95\% uncertainty interval of \textit{posterior median} return values for the segment length given on the x-axis. This empirical estimate of uncertainty is obtained by reading off the 2.5 and 97.5 percentiles of the set of median return values over all starting years for a given segment length. With respect to the right-hand abscissa, the solid, dashed and dotted orange lines indicate the percentage of starting years for which (a) the estimated {posterior median} return value for given segment length lies outside the 95\% credible interval estimated using the full sample, (b) the estimated {posterior median} return value using the full sample lies outside the 95\% credible interval estimated using the given segment length, and (c) there is no overlap at all between 95\% credible intervals estimated from the full sample and from a given segment length. 

For both EoM and SoA, the width of the 95\% credible interval is very large (around 5m) for 20-year segments, reducing with increasing segment length as might be expected to around 1m. For small segment lengths, the chance that the estimated return value lies outside the 95\% credible interval based on the full sample analysis is around 70\%, reducing with increasing segment length to around 30\% for a segment length of 165 years. Conversely, the chance that the (posterior median) return value estimated using the full sample lies outside the 95\% credible interval estimated for a particular segment length is small, around the expected 5\% level. Supporting plots are provided in Figures SM5.20-22. We might surmise from the left panel of Figure~\ref{Fgr:StdStt-CvrEoM} that there might still be some influence of a mixed distribution of storms increasing observed uncertainties in 100-year $H_S$ EoM.


\subsection*{Non-stationary EV analysis}
\begin{figure}[h!]
	\centering
	\includegraphics[width=0.70\columnwidth]{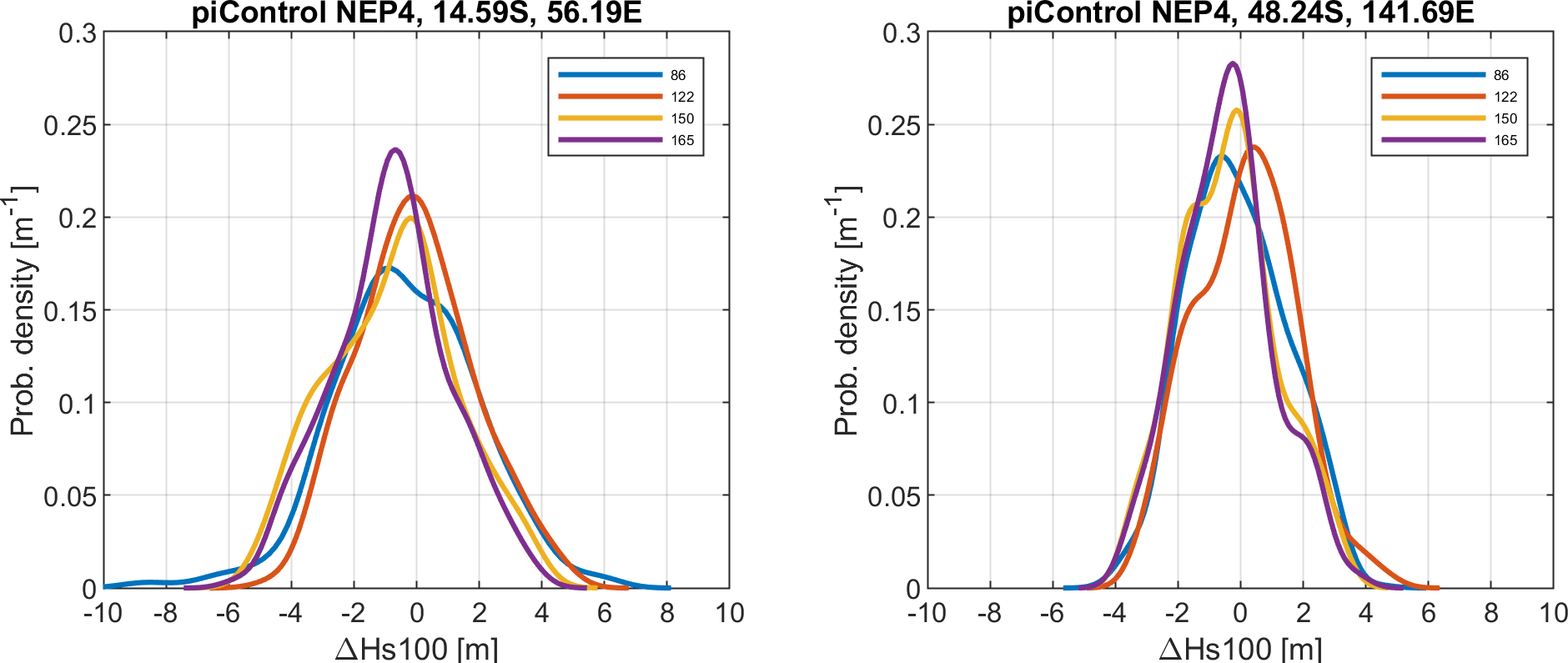}
	\caption{Kernel density estimates of the median difference in 100-year POT $H_S$ (over the length of the segment) estimated using an arbitrary segment of CMIP6 piControl data of given length (in years, see panel legend, from a total run length of 700 years) and a non-stationary GP model with EV threshold level NEP4. Estimates centre location EoM (left) and SoA (right).}
	\label{Fgr:StdStt-NsEva-Pdf}
\end{figure}

It is interesting also to assess the variability in estimates for 100-year POT $H_S$ from \textit{non-stationary} analysis of the piControl output. This gives a direct quantification of the change in return value that might be expected, estimated from samples corresponding to a specific time interval of data, due entirely to inherent steady-state climate variation, when applying non-stationary EV models. It is then particularly interesting to compare the size of this inherent variability in return value, with that associated with different forcing scenarios in Section~\ref{Sct:TmpTrn}, using the same non-stationary EV model forms. Figure~\ref{Fgr:StdStt-NsEva-Pdf} shows estimates for the probability density function of the median difference in 100-year $H_S$ (over the segment length) obtained from non-stationary GP analysis of a segment sample of given length (in years) in the piControl output. We estimate the density as a model average over posterior median difference estimates from non-stationary EV models for all available unique segments of a given length in the piControl output. The choice of segment lengths considered again reflects the lengths of CMIP5 and CMIP6 output available for different climate scenarios (see Section~\ref{Sct:Dat}). The segment length 122 years is a special case, corresponding to the total time span of CMIP5 output. For this segment length, to mimic the time-structure of the CMIP5 data, we subsample the 122 years of piControl output into three appropriately-spaced intervals of lengths 27, 20 and 20 years; therefore the actual number of years of data available for analysis in this case is 67. 

We see that density estimates are relatively unaffected by choice of segment length between 86 and 165 years, for centre locations EoM (left) and SoA (right). Densities are approximately symmetric, and centred around zero. The density EoM is somewhat broader than SoA. At either location, were we to choose segments of data of length between 86 and 165 years for analysis at random, and use them to estimate the change in median 100-year POT $H_S$ over the segment length, we would not be particularly surprised to observe variation of $\pm$ 3m regardless of the segment length. Figures SM5.24-31 and SM5.32-39, provide supporting visualisation of the time evolution of the posterior distributions of 100-year POT $H_S$ at the start and end years. Figures SM5.40-47 illustrate density estimates for the median difference in 100-year POT $H_S$ for all individual combinations of location and segment length.

\FloatBarrier
\section{Temporal trends} \label{Sct:TmpTrn}
%
We now turn to quantifying evidence supporting a temporal trend in return value for the EoM and SoA locations, assessed by fitting a non-stationary EV model (see Section~\ref{Sct:Eva}) to POT or BM data corresponding to a particular choice of GCM and forcing scenario. The fitted model provides an estimate for the joint distribution of return value for all years of interest in the periods 1979-2100 (CMIP5) and 2015-2100 (CIMP6) and beyond. Derived quantities of interest, e.g. the probability that the return value will increase over some period of time, or the expected size of the change in the return value over time, can then be estimated. Finally, we can average our summary statistics over GCMs to obtain ``best overall estimates'' for the quantities of interest under a given forcing scenario. 

We estimate the non-stationary GP model introduced in Section~\ref{Sct:Eva}, for POT of all CMIP5 and CMIP6 output for all scenarios. For each combination of GCM and climate scenario, the MCMC analysis provides a sample from the joint posterior distribution of model parameters. We are then able to calculate the corresponding distribution of other quantities of interest, such as the 100-year POT $H_S$ at any time. In particular, we are able estimate the probability that the value of 100-year POT $H_S$ for the final year (referred to as ``Hs100End'', for the year 2100) is greater than the corresponding value in the first year (referred to as ``Hs100Start'', taken here to be the first year of CMIP5 data, 1979). Note that this requires extrapolation in time under the fitted EV model for CMIP6 data.

\begin{figure}[h!]
	\centering
	\includegraphics[width=0.70\columnwidth]{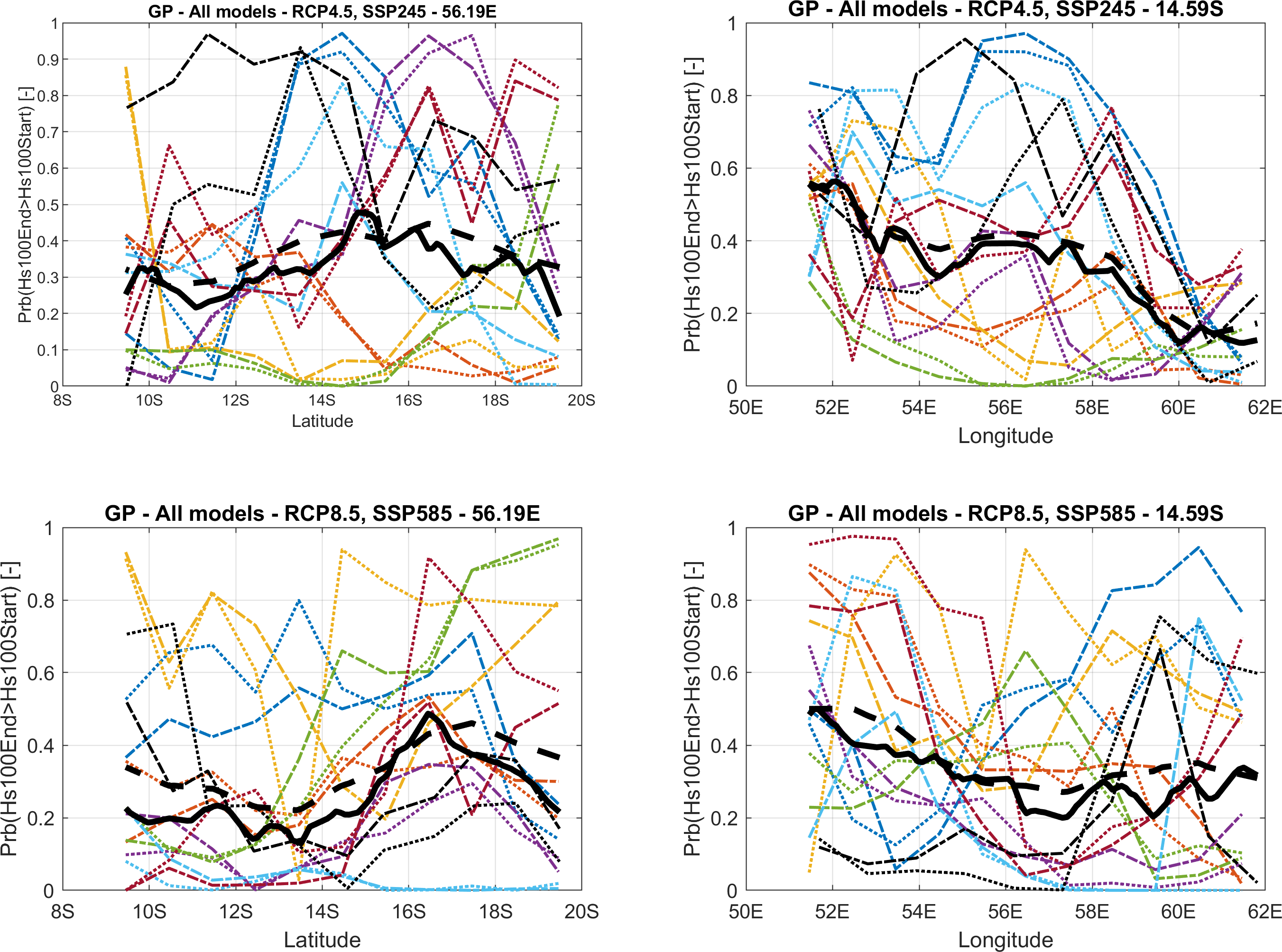}\\
	\caption{Summary of RCP trends for 100-year POT $H_S$ and all GCMs EoM. Panels show the meridional (left) and zonal trends observed for RCP4.5 (and SSP245, top) and RCP8.5 (and SSP5-85, bottom) output. Thin coloured lines represent return value using EV threshold NEP3 (dash-dot) and NEP4 (dot) for individual CMIP5 and CMIP6 GCMs, and the thick black lines provide the global mean (solid) and the global median (dashed) over all GCMs and NEPs. The colour scheme for CMIP5 GCMs is given in Figures~\ref{Fgr:CmpPotTalEoM} and \ref{Fgr:CmpPotRatSoA}; in addition, thin black lines are used to represent the CMIP6 values.}
	\label{Fgr:TmpTrn-RCPSmm-EoM}
\end{figure}

\begin{figure}[h!]
	\centering
	\includegraphics[width=0.70\columnwidth]{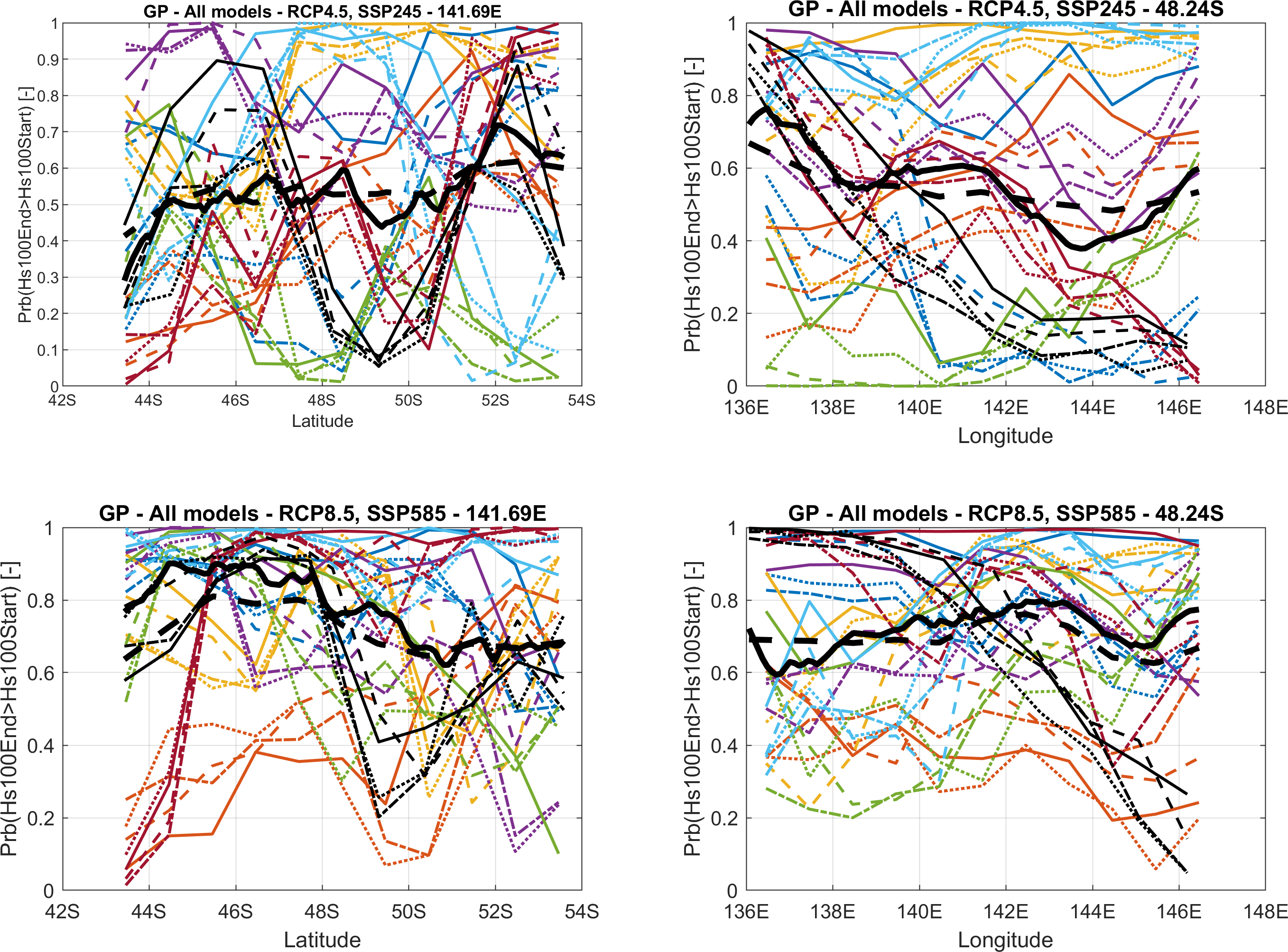}\\
	\caption{Summary of RCP trends for 100-year POT $H_S$ and all GCMs SoA. Panels show the meridional (left) and zonal trends observed for RCP4.5 (top) and RCP8.5 (bottom) output. Thin coloured lines represent return value using EV threshold NEP1 (solid), NEP2 (dashed), NEP3 (dash-dot) and NEP4 (dot) for individual CMIP5 and CMIP6 GCMs, and the thick black lines provide the global mean (solid) and the global median (dashed) over all GCMs and NEPs. The colour scheme for CMIP5 GCMs is given in Figures~\ref{Fgr:CmpPotTalEoM} and \ref{Fgr:CmpPotRatSoA}; in addition, thin black lines are used to represent the CMIP6 values.}
	\label{Fgr:TmpTrn-RCPSmm-SoA}
\end{figure}

Figures~\ref{Fgr:TmpTrn-RCPSmm-EoM} and \ref{Fgr:TmpTrn-RCPSmm-SoA} summarise our findings for RCP4.5 (and SSP245) and RCP8.5 (and SSP585) for EoM and SOA. Panels show the meridional (left) and zonal (right) trends observed for RCP4.5 (top) and RCP8.5 (bottom). Thin coloured lines represent 100-year POT $H_S$ for individual CMIP5 and CMIP6 GCMs, with different line styles for each threshold level NEP1-4 (with only NEP3 and NEP4 for EoM). The thick black lines provide the global mean (solid) and the global median (dashed) over all GCMs and NEPs. There is huge variability between estimated probabilities for different GCMs, and large meridional and zonal variability. However, the global mean and median show more consistency. For EoM, global means and median probabilities tend to be less than 0.5 (suggesting a decrease in return value), whereas for SoA the global probability estimates tend to be larger than 0.5. For SoA, we also see that global estimates of probability for RCP8.5 are larger than for RCP4.5.

As supporting information, Figures SM6.1-8 illustrate meridional and zonal trends in the probability of increased return value for individual GCMs, and Figures SM6.9-16 illustrate the corresponding trends for non-stationary GEV analysis using AM; the use of 5YM data is not possible because of the limited sample length available, and results for EoM should be treated with caution as a result. Figures SM6.1-8 suggest there is large variation in estimates for the probability of a change in return value between GCMs, and with both latitude and longitude, for RCP4.5 and RCP8.5 scenarios. EoM, the few occurrences of a ``significant'' probability (indicated by coloured dots at the locations concerned; see SM6 for further discussion) typically suggest a decreasing trend in return value, although there are exceptions. The opposite is true SoA. Figures SM6.9-16 show broadly similar trends for non-stationary EV analysis of AM. Figures SM6.25-28 illustrate results for the extended set of CMIP6 scenarios, based on non-stationary EV analysis for POT. Again, there is general consistency in estimates of the probability of change in return value for appropriate choice of EV threshold level. However, the large variation with longitude for a given forcing scenario in particular seems difficult to explain on physical grounds. SoA, it is interesting that the 1\% CO$_2$ and abrupt 4 $\times$ CO$_2$ scenarios tend to be associated with significant reductions in return value. There is some evidence here again for erratic behaviour of EoM estimates using NEP3 (e.g. Figure SM6.25-26 for abrupt 4 $\times$ CO$_2$), possibly suggesting the presence of a mixed population of storms \textit{under certain scenarios only} at this threshold level. Figures~SM6.29-32 show corresponding plots for analysis of CMIP6 AM; unfortunately, again because of sample length, analysis of 5YM is not feasible.

\begin{figure}[h!]
	\centering
	\includegraphics[width=0.70\columnwidth]{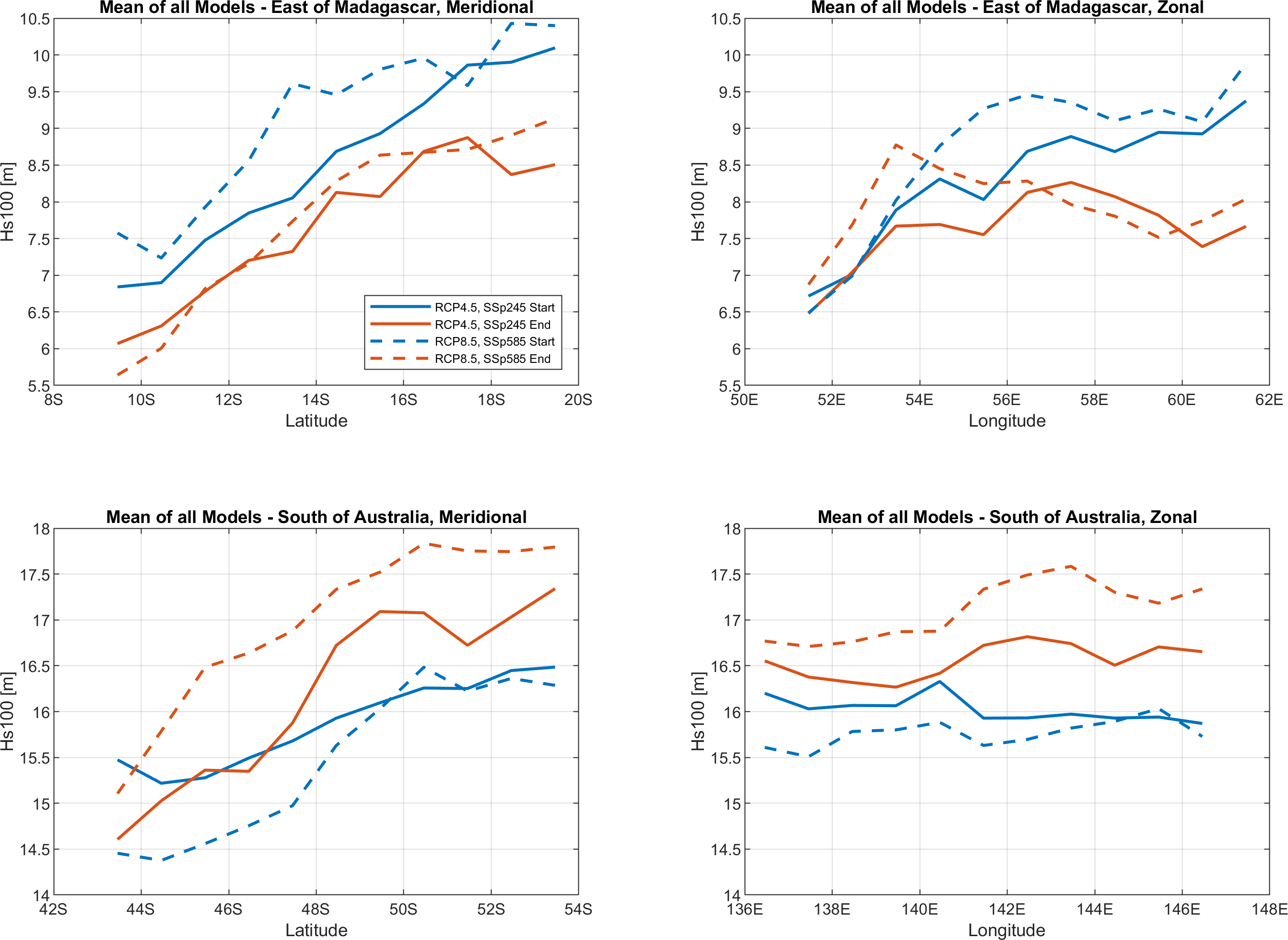}\\
	\caption{Summary of inferences for 100-year POT $H_S$ return value from all CMIP5 and CMIP6 GCMs. Panels show the mean (over all GCMs) of the posterior median return value at the start time (blue) and the end time (orange), for RCP4.5 (and SSP245, solid) and RCP8.5 (and SSP585, dashed) for EoM (top) and SoA (bottom) meridionally (left) and zonally (right).}
	\label{Fgr:TmpTrn-GlbSmm}
\end{figure}

We summarise our findings regarding the values of 100-year POT $H_S$ at the start (1979) and end (2100) of the observation period in Figure~\ref{Fgr:TmpTrn-GlbSmm}. The figure summarises our beliefs about the return value at the start and end time points in terms of the mean (over all GCMs) of the posterior median return value (per GCM) meridionally and zonally, for both EoM and SoA regions and each of the RCP4.5 (and SSP245) and RCP8.5 (and SSP585) scenarios. For both regions, there is evidence that return value increases with increasing latitude, but there is little zonal variation. For the EoM region, there is general evidence for both RCP scenarios that return value reduces in time by approximately 0.5m. SoA, there is general evidence for an increase in return value over the period of approximately 0.5m for RCP4.5 and 1.0m for RCP8.5. 
Supporting Figures SM6.17-24 give estimates for the change in 100-year POT $H_S$ for individual GCMs, and estimate the significance of those changes, under the RCP4.5 and RCP8.5 scenarios. Results again show large variability across GCMs, and little evidence of significant changes in return value. However, there is a tendency for generally reducing trends EoM, and generally neutral or increasing trends SoA. Figures SM6.33-44 give a detailed breakdown of estimates for the meridional and zonal variation of return value for individual GCMs and the start and end years of the study.

\FloatBarrier
\section{Discussion and conclusions} \label{Sct:DscCnc}
%
Estimation of return values from relatively small samples of data is problematic (e.g. \citealt{Srn15}, \citealt{JntEA20}). In their basic analysis step for CMIP5 output, \cite{MccEA20} use a stationary EV analysis of data for a fixed time period of approximately 20 years. They then compare return value estimates, and report that, regardless of which individual GCM is considered, only rarely is a statistically significant change (at 95\% level) in 100-year $H_S$ at a location estimated. For this reason, those authors then aggregate standardised data from different GCMs to increase sample size for analysis (but again, for 20-year time periods only). Using the resulting composite data, significant changes in 100-year $H_S$ are found. In the current analysis, we choose to estimate independent non-stationary EV models for the output of each GCM, using all available data (for CMIP5 or CMIP6) in one analysis, and then estimate the statistical characteristics of return value, and time differences in return value per GCM. Finally, we combine inferences about return values and return value differences over GCMs. We believe our approach has advantages, for the following reason: \cite{MccEA20} combines samples of data drawn from different GCMs, likely to have different statistical properties, together and assumes that extremes of the composite sample can be considered drawn from a homogeneous tail. Although the standardisation step used by \cite{MccEA20}, equating the first and second sample moments of the distributions from each GCM is widely used, the resulting composite sample corresponds to a mixture of different distributions with different tail characteristics in general. Estimating a common tail using such data usually leads to bias, as shown in \cite{JntEwnFrr08a}. The approach used in the current work treats each GCM as an independent source of estimates for 100-year $H_S$ and its uncertainty using a statistical model appropriate for all relevant data. We then come to a final view by aggregating 100-year $H_S$ estimates over all GCMs. Despite the differences in approach, the overall conclusions of this work are broadly in agreement with those of \cite{MccEA20}.

Future climate change projections are often derived from ensembles of simulations from multiple GCMs using weighting schemes. In this work, we apply EV analysis to each of a number of GCMs independently, and then adopt a ``one model, one vote'' averaging to calculate our best estimates. Other studies propose more sophisticated weighting; the statistical principles supporting such weighting schemes are well-established (e.g. \citealt{VhtOjn12}). Indeed, software such as GCMeval (\citealt{PrdEA20}) goes some way to automating the selection of GCMs providing good performance in predicting temperature and precipitation. \cite{McSJns13} demonstrate that spatial smoothing of model projections can provide more informative estimates for a neighbourhood than estimates for single locations in the neighbourhood.  \cite{SnsEA13} present a study of CMIP5 North Atlantic storm track data, in which they find that cyclone frequency climate change response is not found to be model dependent over most of the region, despite large variations in historical storm tracks. \cite{Vnm17b} reports that spatial smoothing via regional frequency analysis can be a useful tool to estimate climate change more efficiently for storms in the North Atlantic.

In the current work, we observe considerable variability in predictions of changes in storm severity at any of a number of locations east of Madagascar (EoM) and south of Australia (SoA) from a set of 7 CMIP5 and one CMIP6 GCM. Examination of predicted storm severity along meridional and zonal transects within each region also reveals large differences in return value estimated at a given year, and change in return value over a number of years. It is reasonable therefore to be rather suspicious that specific inferences from the work (e.g. for specific GCMs, locations, years or choices of modelling hyper-parameters) may be biased to a considerable extent, compared to more general inferences (e.g. those summarising over GCMs, locations, years and modelling hyper-parameters).

We adopt non-stationary extreme value (EV) models for estimation of return values, within which all model parameters are allowed to vary linearly with time. We consider the restriction to linear trends appropriate, since we expect that the effect of climate change on the tail of the distribution of $H_S$ to be relatively small; hence the data for analysis is unlikely to provide evidence in favour of more complex models. We estimate the models using Bayesian inference, using simple MCMC algorithms coded in MATLAB, available at \cite{Jnt21a}. We see that careful specification of the EV model, and choice of EV threshold (for POT analysis) or block length (for BM analysis) is critical. EoM, we find that relatively large thresholds (POT, or block lengths for BM) are necessary to avoid bias in return values causes by a mixed population of extratropical storm and occasional tropical cyclones. SoA, we find that threshold (or block length) selection is less critical. Results using non-stationary GP modelling of POT are generally in very good agreement with those from non-stationary GEV modelling of BM, provided that thresholds and block lengths are set sensibly. Indeed, the estimated return value (and change in return value) appears less sensitive to threshold (or block length) choice than to the choice of GCM, or the arbitrary choice of a location with a geographic neighbourhood, again provided that the range of thresholds (or block lengths) admitted is physically plausible to avoid issues with fitting to heterogeneous tails.

Non-stationary EV analysis of the assumed steady-state 700-year piControl data suggests that estimates as large as $\pm3$m for the change in 100-year return value of POT $H_S$ in $N$ years, made from samples corresponding to $N$ years of observation for $N \in \{86,122,150,165\}$, are not unusual. We propose a simple time-randomisation scheme to assess the significance of observed changes in 100-year $H_S$, and use the 700-year long piControl output from the CMIP6 GCM to explore the sensitivity of estimates of return value to the ``segment'' length of data available as the basis for estimation, and the arbitrary time during the 700-year period at which the data are selected. For a segment length of 20 years for EV analysis, the 95\% credible interval for 100-year $H_S$ is around 5m wide, reducing to around 3m for a segment length of 100 years. For a 20-year segment length, there is a probability of around 0.7 that the estimated 100-year return value \textit{lies outside} the 95\% credible interval for the return value using a segment length of 700 years. Large meridional and zonal uncertainties in return value estimates for small segment lengths in particular suggests that joint spatial modelling may be advantageous, or more pragmatically that making best estimates for a neighbourhood by averaging return values estimates, made independently for locations within the neighbourhood, may be useful (\citealt{McSJns13}). Despite huge between-GCM and large within-neighbourhood variability found, averaging predictions over climate models and space suggests that a lowering of storm severity is likelier than not EoM under RCP4.5 (and SSP245) and RCP8.5 (and SSP585) scenarios, whereas the opposite is true SoA. 

\FloatBarrier
\section*{Acknowledgement} \label{Sct:Ack}
%
The authors would like to thank Simon Brown and Rob Shooter (UK Metoffice), Alberto Meucci and Ian Young (U. Melbourne, Australia) and Zhenya Song (First Institute of Oceanography, Qingdao, China) for discussions. The supplementary material (SM) for the current article is available online at \cite{EwnJnt22}. Software for the non-stationary extreme value analysis is provided at \cite{Jnt21a}.

\appendix
\section{Bayesian inference} \label{Sct:App:BysInf}
%
Inference for the GEV, GP, quantile and Poisson regression models described in Section~\ref{Sct:Eva} is performed using Markov chain Monte Carlo (MCMC, see e.g. \citealt{GmrLps06}) following the method of \cite{RbrRsn09}. In this approach, all the parameters $\boldsymbol{\theta}$ of the model are jointly updated for a sequence of $n_B+n_I$ MCMC iterations. At each iteration, a new set of parameter values is proposed, and accepted according to the Metropolis-Hastings acceptance criterion based on (a) the sample likelihood evaluated at the current and candidate states, and (b) the values of the prior densities for parameters at the current and candidate states. Following a certain number $n_B$ of so-called burn-in iterations, the Markov chain is judged to have converged, so that the subsequent $n_I$ iterations provide a valid sample from the joint posterior distribution of parameters.  

Diffuse prior distributions were specified as $\xi \sim U(-0.5,0.2)$; $\sigma \sim U(0,10)$; $\mu, \psi \sim U(-10, 10)$. Likelihoods for the models are available from the distributions given in the main text. An appropriate starting solution $\boldsymbol{\theta}_1$ for the MCMC inference was obtained by random sampling from the prior distributions of parameters, ensuring a valid likelihood. 

For the first $n_S<n_B$ iterations, candidate parameter values $\boldsymbol{\theta}_k^c$ are proposed (independently) from $\boldsymbol{\theta}_k^c \sim N(\boldsymbol{0},0.1^2 \boldsymbol{I})$ following \cite{RbrRsn09}. Thereafter $\boldsymbol{\theta}_k^c \sim (1-\beta) N\left(\boldsymbol{\theta}_{k-1}, 2.38^{2} \Sigma_{k}\right)+\beta N\left(\boldsymbol{\theta}_{k-1}, 0.1^{2} / 4\right)$, where $\beta=0.05$, $\Sigma_{k}$ is the empirical variance-covariance matrix of parameters from the past $k$ iterations, and $\boldsymbol{\theta}_{k-1}$ is the current value of parameters.

Throughout, a candidate state is accepted using the standard Metropolis-Hasting acceptance criterion. Since prior distributions for parameters are uniform, and proposals symmetric, this criterion is effectively just a likelihood ratio. That is, we accept the candidate state with probability $\min (1, L(\boldsymbol{\theta}^c)/L(\boldsymbol{\theta}))$, where $L(\boldsymbol{\theta})$ and $L(\boldsymbol{\theta}^c)$ are the likelihoods evaluated at the current and candidate states respectively, with candidates lying outside their prior domains rejected.

\section{Propagation of GP threshold uncertainty} \label{Sct:App:GpUnc}
As discussed in Section~\ref{Sct:Eva:Gp}, in this work the posterior mean quantile regression thresholds $\hat{\psi}^S$ and $\hat{\psi}^E$ (corresponding to the start and end years of data) are used for subsequent estimation of EV and threshold exceedance rate models. Posterior mean $\hat{\psi}^S$ and $\hat{\psi}^E$ are also used for simulation under the model, and estimation of distributions for $T$-year return values. Since estimation of EV models is generally considerably more uncertain that EV threshold, our neglect of threshold uncertainty was not considered an important deficiency. To investigate this concern further, we estimated the distribution of $100$-year return value for the start and end years using fitted models, under different procedures for propagation of uncertainty in EV threshold. Results are illustrated in Figure SMB.1 for a sample corresponding to an interval of $P=86$ years of PI Control data offshore Madagascar.

Three possible approaches to propagating threshold uncertainty are illustrated, and correspond to different choices regarding the non-stationary threshold estimate used for simulation (see Equation~\ref{E:CvrTim}, with $\eta=\psi$) based on $\psi^S$ and $\psi^E$. These choices use (a) independent draws of $\hat{\psi}^S$ and $\hat{\psi}^E$ from their posterior marginal distributions, (b) a common draw of $(\hat{\psi}^S, \hat{\psi}^E)$ from its posterior joint distribution, and (c) the posterior mean values of $\hat{\psi}^S$ and $\hat{\psi}^E$. The figure shows that variation in the distribution of the 100-year return value from choices (a)-(c) is small, for the start and end years, and for the different threshold non-exceedance probability values $\tau$ given in the four panels. Similar results are observed for other locations and time periods. For this reason, we consider it reasonable to use the posterior mean $\hat{\psi}^S$ and $\hat{\psi}^E$ for estimation of return values.

\bibliographystyle{elsarticle-harv}
\bibliography{C:/Users/Philip.Jonathan/PhilipGit/Code/LaTeX/phil}

\end{document}